\documentclass[12pt]{iopart}
\usepackage[title]{appendix}
\usepackage{iopams}
\usepackage{graphicx}
 \usepackage[colorlinks, citecolor = blue, urlcolor = blue]{hyperref}
 \usepackage{breakurl}

\begin{document}

 
\title[On the subtleties of statistical isotropy in cosmic topology]
{On the subtleties of statistical isotropy in cosmic topology}
\author{Ralf Aurich$^1$, Martin J France$^2$ and Frank Steiner$^1$}
 
\address{$^1$ Institut f\"ur Theoretische Physik, Universit\"at Ulm,\\
  Albert-Einstein-Allee 11, D-89081 Ulm, Germany\\
$^2$ Universit\'e Lyon 1, ENS de Lyon, CNRS, CRAL, UMR 5574, Saint-Genis-Laval, France}

\begin{abstract}
The cosmological principle assumes homogeneity and isotropy on the
largest scales for which the cosmic microwave background (CMB) provides
the best testing grounds.
Here, an analysis of the statistical isotropy for the CMB in the case of
the cubic 3-torus topology  $\mathbb{T}^3$ is presented.
Since the statistical anisotropy in the CMB 3-torus sky maps is very tiny,
the analysis requires very large ensembles consisting of 100\,000 CMB
simulations in order to reveal deviations from the isotropy.
Orbit sum rules, which also have applications in quantum chaos,
are used to describe the anisotropy structures found in the torus ensembles.
They relate the length of orbits, which connect a given point on the surface
of last scattering with the topological mirror points of the CMB observer,
to the anisotropy pattern.
It is found that the positions on the surface of last scattering
which are closest to mirror observers possess the largest deviations from
the statistical isotropy.
Furthermore, the anisotropy of the normalised modulus of the gradient of
the CMB temperature field is discussed.
\end{abstract}





\section{Introduction}

The cosmic microwave background (CMB) radiation is almost perfectly
statistically isotropic and is one of the strongest pillars of empirical
evidence for the cosmological principle
which assumes homogeneity and isotropy on the largest scales.
In order to achieve this important result, the monopole and the kinematic
dipole of the CMB as well as foreground emission have to be subtracted.
Since the availability of full-sky CMB surveys,
cosmologists ponder the question of a violation of the statistical isotropy.
The most recent data are analysed by the Planck team in
\cite{Planck_2018_VII} with respect to the statistical isotropy.
A good survey of the diagnostic tools to detect a violation of the
statistical isotropy can be found in
\cite{Schwarz_et_al_2016,Saadeh_et_al_2016,Jones_et_al_2026}.
One prominent example is the parity asymmetry and the related power asymmetry
between even and odd low-multipoles \cite{Kim_Naselsky_2010}.
The parity asymmetry is further investigated in a lot of papers,
for example see references in \cite{Kester_Bernui_2024} and
points to a violation.
An analysis of the asymmetry in the Planck data using the Minkowski
functionals can be found in \cite{Duque_Barandiaran_Martínez-Arrizabalaga_2026}.
But there are other measures as the correlation matrix
$C_{lml'm'} :=\left< a_{lm} a_{l'm'} \right>$
of the spherical expansion coefficients
$a_{lm}$ of the CMB temperature fluctuations,
which should be diagonal in the case of statistical isotropy,
see e.\,g.\ \cite{Aurich_Janzer_Lustig_Steiner_2007}.

Among the many possibilities for explaining a violation in statistical isotropy
are multiply connected manifolds studied in cosmic topology
\cite{Lachieze-Rey_Luminet_1995,Levin_2002,Hitchman_2008,Copi_et_al_2026},
that inherently break this statistical isotropy without the need for
modified local physics.
This publication deals with the simplest non-trivial topology in the
Euclidean space, the 3-torus.
This model serves as a testing ground for various diagnostic tools
concerning statistical isotropy.
For example, \cite{Hajian_Souradeep_2003} introduces the bipolar power spectrum
which measures how pairs of different angular scales correlate and thus detects
subtle directional patterns.
Furthermore, the expected correlation matrix $C_{lml'm'}$ is now calculated
for all Euclidean topologies in \cite{Eskilt_et_al_2023} for orientable manifolds
and in \cite{Copi_et_al_2025} for non-orientable manifolds.
For the Euclidean orientable case, also the correlation matrix
is known for tensor modes \cite{Samandar_2024,Samandar_2025}.

A more direct tool to decipher the transformations defining non-trivial
topological spaces is given by the circles-in-the-sky test
\cite{Cornish_Spergel_Starkman_1998b}.
Since the CMB radiation map betrays no matched circle pairs,
various topological spaces are excluded,
but see
\cite{Akrami_et_al_2022,Petersen_et_al_2023,Saha_et_al_2025,%
  Noltmann_et_al_2026,Tamosiunas_et_al_2024,Barandiaran_et_al_2026}
for an analysis of spaces
which could escape the circles-in-the-sky test.

This publication is organised as follows.
In section \ref{Signatures_of_isotropy} the variance $\sigma^2(\hat n)$
of the CMB temperature field and the variance $\sigma_1^2(\hat n)$ obtained
from its covariant derivative used in our study of the anisotropy are defined.
These isotropy measures are applied to the cubic 3-torus ensemble
which is defined in section \ref{torus_topology}.
Since the statistical anisotropy proves to be so tiny,
torus ensembles consisting of 100\,000 members are analysed in
section \ref{Anisotropy_for_T3} for $\sigma^2(\hat n)$
and $\sigma_1^2(\hat n)$.
In order to understand the behaviour observed in
section \ref{Anisotropy_for_T3},
we apply orbit sum rules to the ensemble averaging.
In the first step, the Sachs-Wolfe approximation is adapted in
section \ref{Sachs-Wolfe} by using some simplifying assumptions,
while in the next step presented in section \ref{orbit-sum-rule}
the orbit sum rules are applied,
which show that the lengths of the orbits, which connect a point
on the surface of last scattering with the topological mirror points
of the CMB observer,
determine the anisotropy structures.
As an improvement to the crude approximations used
in section \ref{orbit-sum-rule},
section \ref{NSW-with-almost-Harrison-Zeldovitch} presents an analysis
with an almost Harrison-Zeldovitch power spectrum.
In section \ref{gradient-temperature-field} we discuss the influence of
the anisotropy on the normalised modulus of the gradient of
the CMB temperature field which is analysed in
\cite{Aurich_France_Steiner_2026} with respect to different mask orientations.
There it turns out that a faint dependence on the orientation exists and
section \ref{gradient-temperature-field} provides an explanation
for this behaviour.
Finally, section \ref{summary} summarises the results.
In the Appendix the orbit sum rules are derived.


\section{Signatures for the detection of violation of statistical isotropy}

\label{Signatures_of_isotropy}

The CMB temperature field $T(\hat n)$ is now measured with high accuracy
so that the CMB temperature fluctuations
\begin{equation}\label{def_cmb_fluctuations}
  \delta T(\hat n) \; := \; T(\hat n) \, - \, T_0
\end{equation}
around the monopole $T_0$ are the topic of many studies.
Here, $\hat n = \hat n(\vartheta,\varphi)$ denotes the unit vector pointing
towards the direction defined by the spherical coordinates
$(\vartheta,\varphi)$.
The definition (\ref{def_cmb_fluctuations}) implies that the spherical average
\begin{equation}\label{spherical_average}
  \frac 1{4\pi}\int_{{\cal S}^2} d^2\hat n \; \delta T(\hat n) \; = \; 0
\end{equation}
vanishes.
This is the average of $\delta T(\hat n)$ with respect to a single
CMB sky realisation,
which can be given by the observed CMB sky or by a single realisation drawn
from an ensemble of CMB realisations belonging to a given set of
cosmological parameters.
With respect to an ensemble average, a {\it position dependent} average
\begin{equation}
  \mu(\hat n) \; := \;
  \left< \delta T(\hat n) \right>
\end{equation}
can be obtained
if the assumption of statistical isotropy is violated.
In the following, the brackets $\left< \dots \right>$ always denote
the ensemble average at a fixed direction $\hat n$.
Note, that it trivially follows from (\ref{spherical_average})
that the spherical average of $\mu(\hat n)$ nevertheless vanishes
\begin{equation}\label{spherical_average_mu}
  \frac 1{4\pi}\int_{{\cal S}^2} d^2\hat n \; \mu(\hat n) \; = \; 0
  \hspace{10pt} .
\end{equation}
A possible position dependence as a signature of the violation of the
statistical isotropy is the central topic of this paper.

The ensemble average of the variance of $\delta T(\hat n)$ is defined as
\begin{equation}\label{def_variance}
  \sigma^2(\hat n) \; = \;
  \left<\, \left( \delta T(\hat n) - \mu(\hat n) \right)^2\, \right>
  \hspace{10pt} .
\end{equation}
The standard deviation $\sigma(\hat n)$ being simply the square root of
the variance (\ref{def_variance}) is the central quantity of our analysis.
Other quantities are also investigated but it is found
that the standard deviation $\sigma(\hat n)$ reveals the strongest signature.

Nevertheless, an example of a further signature for possible
anisotropy detection is provided by the variance $\sigma_1^2(\hat n)$
of the squared modulus of the temperature gradient field.
This can be constructed from the covariant derivatives
$\delta T_{,i}(\hat n)$, $i\in\{\vartheta,\varphi\}$, as
\begin{equation}
\label{def_sigma_1}
\sigma_1^2(\hat n) \, := \,
\left<\delta T^{,i}(\hat n)\, \delta T_{,i}(\hat n)\, \right>
\, = \,
\left< \left(\frac{\partial\delta T(\hat n)}{\partial\vartheta}\right)^2
\, + \,
\left(\frac 1{\sin\vartheta}\,\frac{\partial\delta T(\hat n)}{\partial\varphi}\right)^2
\right>
\hspace{4pt} .
\end{equation}


\section{The cubic 3-torus topology ${\mathbb{T}}^3$}

\label{torus_topology}


The cosmological standard model, the $\Lambda$CDM model, is based on
an infinite flat space.
Therefore, a flat space form is considered here,
although there exist examples also for the spherical and the hyperbolic spaces.
Flat space forms are constructed by the quotient ${\mathbb E}^3/\Gamma$ of
the three-dimensional Euclidean space ${\mathbb E}^3$ by a discrete symmetry
group $\Gamma$ which is free of fixed points.
There are 18 topologically distinct space forms in ${\mathbb E}^3$.
Restricting to space forms with finite volume excludes eight of them,
and ignoring non-orientable manifolds excludes further four flat space forms,
so that there are six flat orientable space forms with finite volume
\cite{Lachieze-Rey_Luminet_1995,Riazuelo_et_al_2004}.
From these six models, the only globally homogeneous space form is chosen,
which is also the simplest one: the 3-torus ${\mathbb{T}}^3$.

The symmetry group $\Gamma$ of a general 3-torus ${\mathbb{T}}^3$ is generated
by three translations in the direction of three linearly independent,
oriented vectors $\vec a_i (i = 1, 2, 3)$ of length $L_i$
which meet at a common point.
These vectors define the edges of a general parallelepiped
which is identified as the fundamental domain of a general 3-torus.
The parallelepiped has 12 edges, 8 vertices and 6 faces (parallelograms),
whose opposite faces are identified by pairs.
If the vectors $\vec{a_i}$ are pointing in the direction of the unit vectors
$\hat{e_i} (i = x, y, z)$,
the fundamental domain is a cuboid (rectangular box).
In this paper, we consider the special case of a cubic 3-torus for which all 
lengths are equal, $L_x = L_y = L_z = L$.
(In the following the torus length $L$ is given in units of the
Hubble length $L_H=c/H_0$.)
It should be noted that all 3-tori are globally anisotropic,
as the periodic boundary conditions break the rotational SO(3)-symmetry.

In order to analyse the anisotropy properties of the CMB of this
globally homogeneous, but nevertheless globally anisotropic space form,
one has to expand the perturbations of the various components of the
$\Lambda$CDM model with respect to the eigenfunctions of the
Laplace operator on the considered space form.
The normalised eigenfunctions $\psi_{\vec \alpha}(\vec x\,)$ of the
cubic 3-torus ${\mathbb{T}}^3$ are given by
\begin{equation}\label{eigenfunction}
  \psi_{\vec \alpha}(\vec x\,) \; = \;
  \frac 1{\sqrt{V}}\,e^{i \vec k_{\vec\alpha}\cdot\vec x}
  \hspace{10pt} , \hspace{10pt}
  V = L^3
  \hspace{10pt} , \hspace{10pt}
  \vec k_{\vec\alpha} = \frac{2\pi}L\,\vec\alpha
  \hspace{10pt} ,
\end{equation}
and $\vec\alpha = (\alpha_1,\alpha_2,\alpha_3) \in \mathbb{Z}^3$.
Furthermore, the eigenvalues are
$E_{\vec\alpha} = \vec k_{\vec\alpha}^2 =
\left(\frac{2\pi}L\right)^2\,\vec\alpha\,^2$.
An important point for the following analysis,
where an ensemble of ${\mathbb{T}}^3$ universes for a fixed value of $L$
is generated, is the fact
that all members of a ${\mathbb{T}}^3$ ensemble are oriented in the same way,
because the same vectors $\vec{a_i}$ are used.
The $x$-$y$-plane is the galactic equator in the following Mollweide projections
of the sky maps.
The $\Lambda$CDM concordance model is specified by the cosmological parameters
as given by the Planck Collaboration in \cite{Planck_2015_XIII}
in their table 4 in the column 'TT+lowP+lensing'.
The main parameters are $\Omega_{\hbox{\scriptsize b}}h^2 =0.02226$,
$\Omega_{\hbox{\scriptsize c}}h^2 = 0.1186$, and $h=0.6781$.
Furthermore, all sky maps are computed with the Healpix resolution
$N_{\hbox{\scriptsize side}}=128$ and a Gaussian smearing of
$120\hbox{ arcmin}$ is used.


\section{Topological violation of statistical isotropy}

\label{Anisotropy_for_T3}


In this section the statistical isotropy is analysed for ensembles of
100\,000 CMB realisations for each torus size defined by its side-length $L$.
Numerically, it turns out that there is no violation of the isotropy
in the temperature fluctuation $\delta T(\hat n)$
so that one gets $\left< \delta T(\hat n) \right> = 0$.
It might be possible that there is a violation but it has then to be so tiny
that it cannot even be extracted on the base of 100\,000 CMB realisations.

\begin{figure}
\begin{center}
\vspace*{0pt}
\begin{minipage}{15.8cm}
\hspace*{0pt}\includegraphics[width=8.0cm,angle=270]{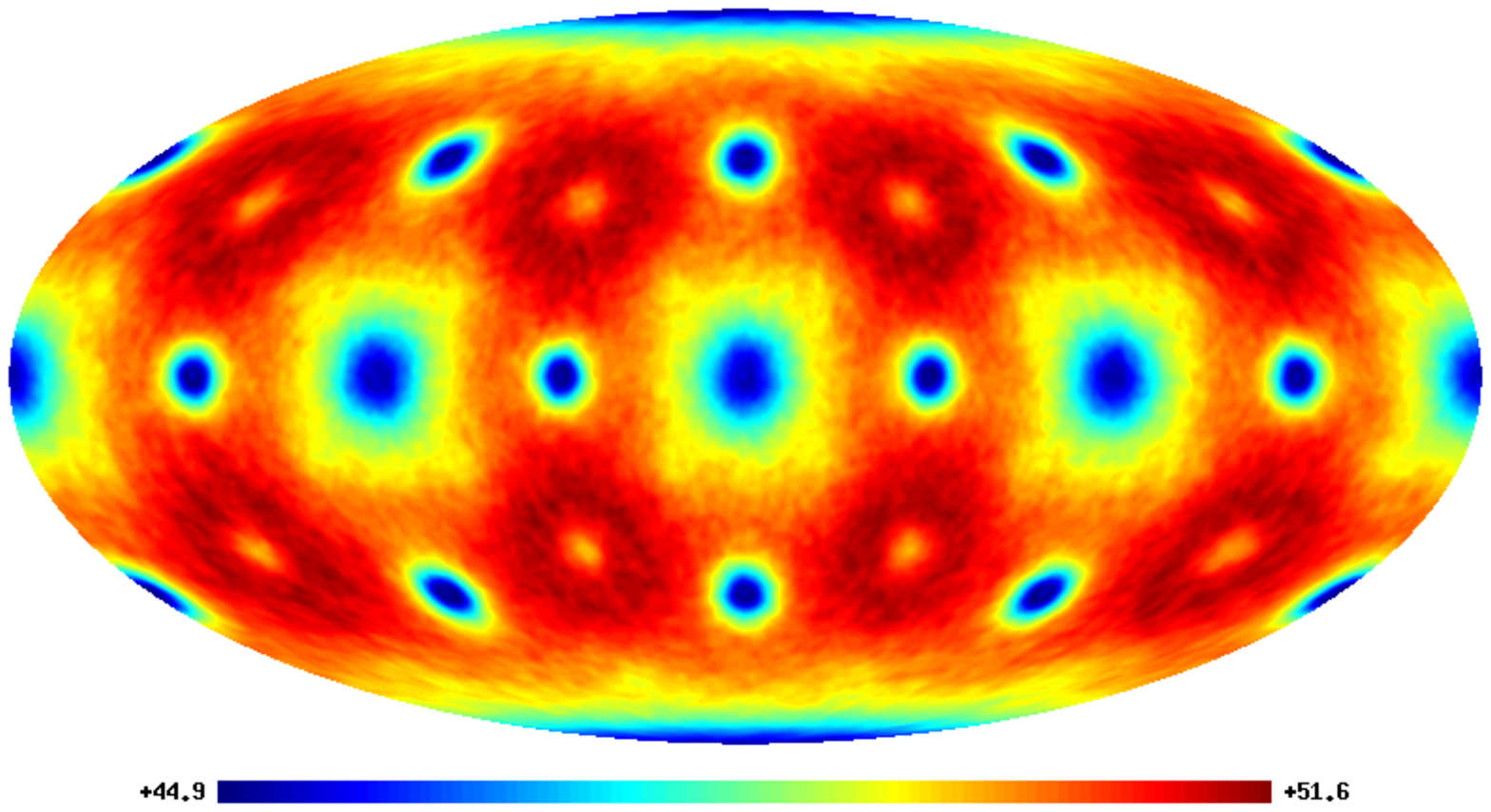}
\end{minipage}
\end{center}
\vspace*{0pt}
\caption{\label{Fig:L2_StdDev}
  The standard deviation $\sigma(\hat n)$
  that is the square root of the variance $\sigma^2(\hat n)$
  defined in eq.(\ref{def_variance}),
  is computed from the $L=2$ ensemble containing 100\,000 realisations.
  The sky map is smoothed with a Gaussian of $\hbox{fwhm} = 2^\circ$.
  The dipole contribution is omitted, as it is the case for observational data.
  The units are $\mu\hbox{K}$.
}
\end{figure}

\begin{figure}
\begin{center}
\vspace*{0pt}
\begin{minipage}{15.8cm}
\hspace*{0pt}\includegraphics[width=8.0cm,angle=270]{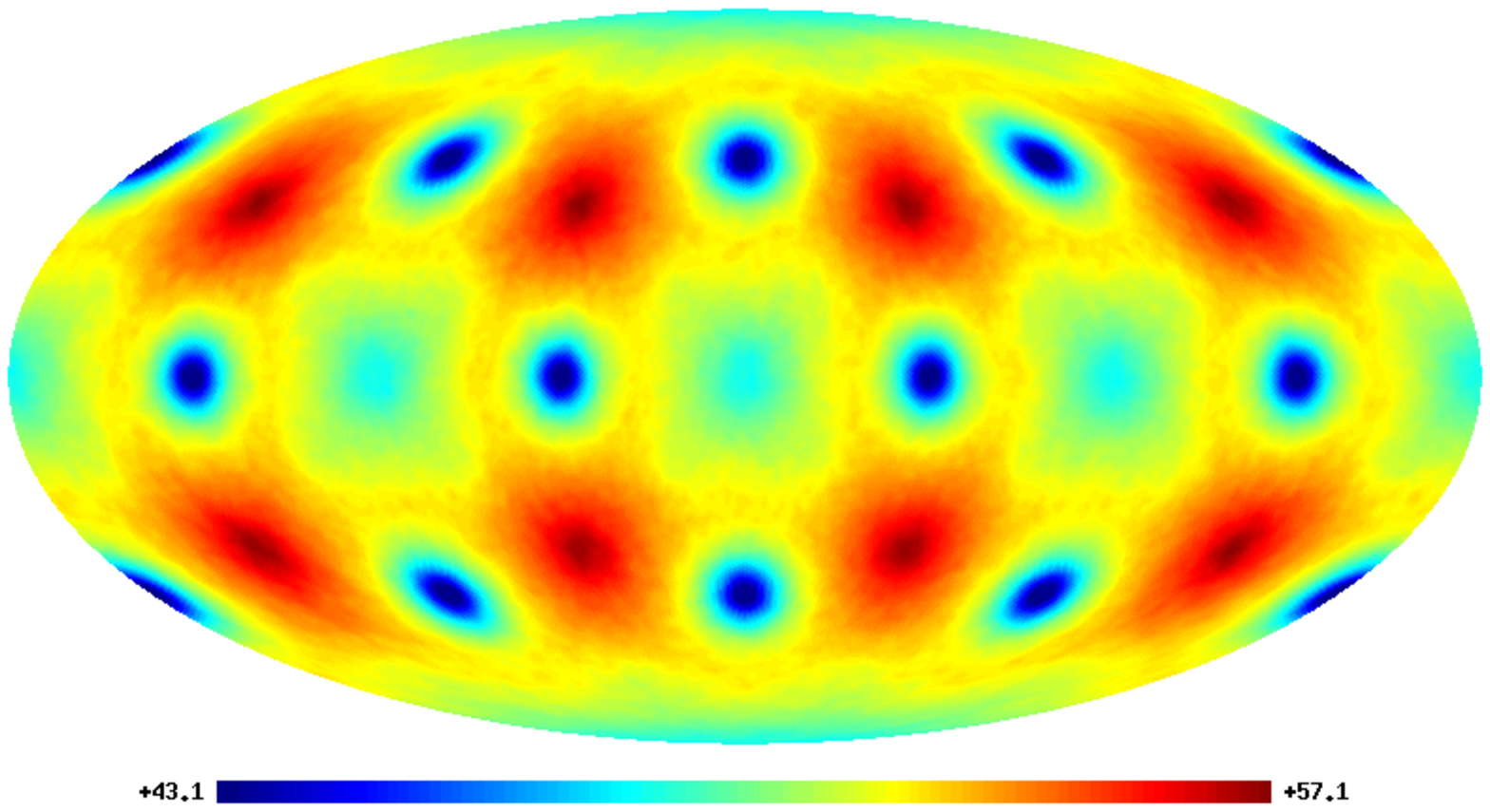}
\end{minipage}
\end{center}
\vspace*{0pt}
\caption{\label{Fig:L2_StdDev_withDip}
  The same as figure \ref{Fig:L2_StdDev}, but now the dipole contribution is
  included.
}
\end{figure}

Thus, the interesting question arises whether higher statistical measures as introduced
in section \ref{Signatures_of_isotropy} are able to reveal a violation of statistical isotropy.
Indeed, the position-dependent standard deviation $\sigma(\hat n)$ being the
square root of the variance $\sigma^2(\hat n)$ defined in eq.(\ref{def_variance})
does show a non-isotropic behaviour.
The example for the ensemble with $L=2$ is shown in figure \ref{Fig:L2_StdDev},
where the dipole contribution is omitted, as it is the case for observational data.
The units are $\mu\hbox{K}$.
The figure \ref{Fig:L2_StdDev_withDip} shows the result without subtracting
the dipole in order to demonstrate the influence of the dipole.

Furthermore, in our analysis, we smooth the sky maps with a Gaussian of
$\hbox{fwhm} = 2^\circ$.
To this aim,
the spherical expansion of the temperature fluctuation $\delta T(\hat n)$
is used
\begin{equation}
  \label{eq:deltaT_spherical_expansion}
  \delta T(\hat n) \; = \;
  \sum_{\ell,m} a_{\ell m} \, Y_{\ell m}(\vartheta,\varphi)
  \hspace{10pt} ,
\end{equation}
where $Y_{\ell m}(\hat n)\equiv Y_{\ell m}(\vartheta,\varphi)$
are the spherical harmonics.
The Gaussian smoothing is carried out in the $a_{\ell m}$ space by multiplying
the expansion coefficients $a_{\ell m}$ with
$F_\ell = \exp(-\frac 12\alpha^2\theta_G^2\,\ell(\ell+1))$ and
$\alpha = \pi/(180\sqrt{8\ln 2})$
if the smoothing $\theta_G$ is given in degrees.
In this way, only the topologically relevant large angular scales are retained.

\begin{figure}
\begin{center}
\vspace*{0pt}
\begin{minipage}{15.8cm}
\hspace*{0pt}\includegraphics[width=8.0cm,angle=270]{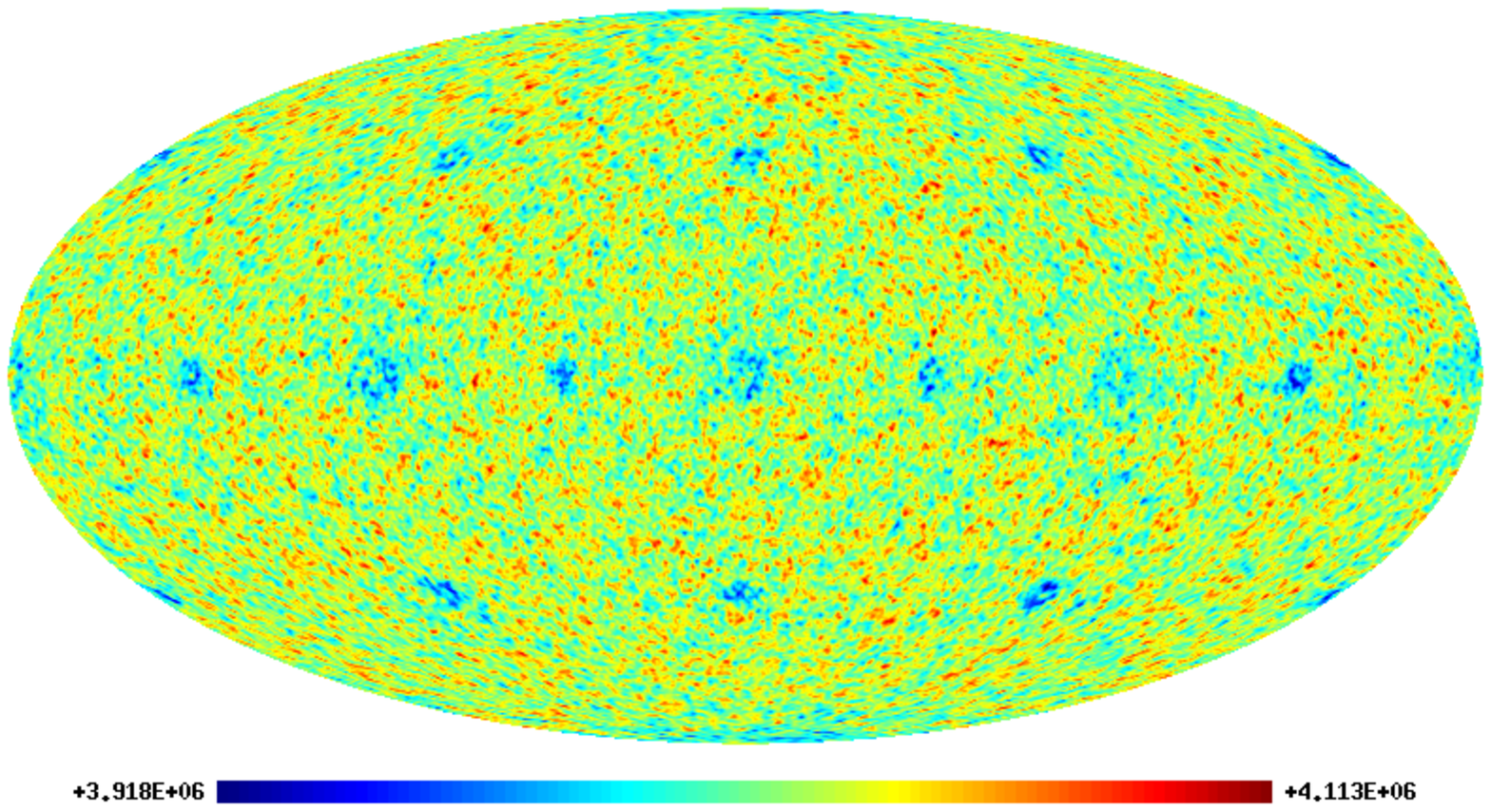}
\end{minipage}
\end{center}
\vspace*{0pt}
\caption{\label{Fig:L2_Square_Derivative}
  The square of the derivative $\sigma_1^2(\hat n)$ defined in
  (\ref{def_sigma_1}) is computed from
  the $L=2$ ensemble containing 100\,000 realizations.
  The sky map is smoothed with a Gaussian of $\hbox{fwhm} = 2^\circ$.
  The units are $\mu\hbox{K}^2$.
}
\end{figure}

In section \ref{Signatures_of_isotropy} the variance $\sigma_1^2(\hat n)$ of
the modulus of the temperature gradient field is introduced
in eq.(\ref{def_sigma_1}).
It turns out numerically that although the anisotropy is visible,
it is much less pronounced compared to the case for $\sigma(\hat n)$.
The result is shown in figure \ref{Fig:L2_Square_Derivative} again for the $L=2$ ensemble.
Only at the extreme points of the standard deviation $\sigma(\hat n)$
seen in figures \ref{Fig:L2_StdDev} and \ref{Fig:L2_StdDev_withDip},
a signature of anisotropy is visible in figure \ref{Fig:L2_Square_Derivative}
although the same smoothing is applied.

\begin{figure}
\begin{center}
\vspace*{0pt}
\begin{minipage}{15.8cm}
\hspace*{100pt}\includegraphics[width=10cm]{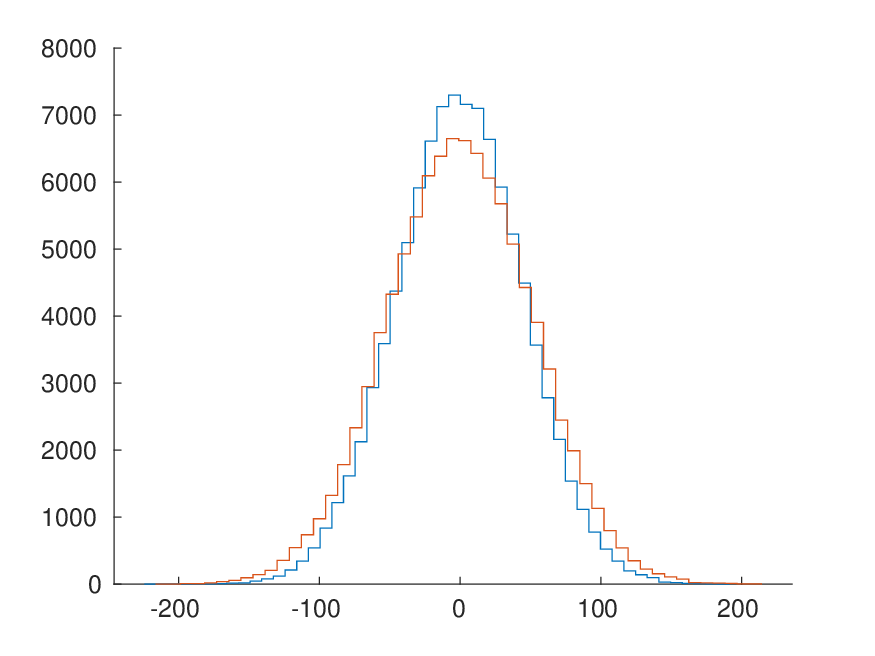}
\end{minipage}
\put(-95,-90){$\sigma$\;\;\; $[\mu\hbox{K}]$}
\end{center}
\vspace*{0pt}
\caption{\label{Fig:Histogram_max_min_comparison}
  The distributions of the standard deviations at two points of the
  standard deviation map $\sigma(\hat n)$ shown in
  figure \ref{Fig:L2_StdDev} are plotted.
  The blue histogram belongs to a point $\hat n$ with a small standard deviation
  of $\sigma=45.238\mu\hbox{K}$, while the red histogram is computed at a position $\hat n$
  having the larger value $\sigma=51.502\mu\hbox{K}$.
  Both histograms are based on 100\,000 realisations.
  One receives from this comparison that the violation of the isotropy
  with respect to the standard deviation $\sigma(\hat n)$ is small
  but nevertheless present.
}
\end{figure}

It is at first sight surprising that the variance $\sigma^2(\hat n)$ violates the isotropy
but not the temperature fluctuation $\delta T(\hat n)$ itself,
so that one obtains numerically $\left< \delta T(\hat n) \right> = 0$.
In order to illustrate this behaviour, figure \ref{Fig:Histogram_max_min_comparison}
shows the histograms derived at two points of the standard deviation map
$\sigma(\hat n)$,
where one point possesses a large standard deviation in the distribution of
$\delta T(\hat n)$ and the other a small one.
One observes in both cases a zero-centred almost Gaussian distribution,
but only the width and the peak height differs.
The blue histogram corresponding to $\sigma=45.238\mu\hbox{K}$
peaks higher than the red one ($\sim 7300$ compared to $\sim 6600$)
which is compensated by wider tails for the latter one.

Because of $\left< \delta T(\hat n) \right> = 0$,
the variance $\sigma^2(\hat n)$ defined in (\ref{def_variance})
can be approximated as
\begin{equation}\label{variance_appr}
\sigma^2(\hat n) \; := \;
\left< \delta T^2(\hat n) \right> \, - \,
\left< \delta T(\hat n) \right>^2 \; \simeq \;
\left< \delta T^2(\hat n) \right>
\hspace{10pt} ,
\end{equation}
which is used in the following discussion.

\section{The Sachs-Wolfe approximation}

\label{Sachs-Wolfe}


The results presented in section \ref{Anisotropy_for_T3}
are based on CMB simulations which include the full Boltzmann physics.
In order to understand the origin of the anisotropy in the CMB maps
of the cubic torus simulations,
it is more instructive to consider the Sachs-Wolfe formula \cite{Sachs_Wolfe_1967}
in the tight-coupling approximation
which allows a connection of the metric perturbation $\Phi(\eta,\vec x\,)$
in terms of the eigenfunctions of the $\mathbb{T}^3$ topology with the
CMB fluctuations.
($\eta$ is the conformal time.)

Before applying the Sachs-Wolfe formula, let us turn to the metric perturbation.
The metric perturbation $\Phi(\eta,\vec x\,)$ can be expanded
with respect to the eigenfunctions (\ref{eigenfunction})
\begin{equation}\label{metric_perturbation}
  \Phi(\eta,\vec x\,) \; = \;
  \sum_{\vec k} \, \Phi_{\vec k}(\eta) \,\psi_{\vec k}(r,\theta,\phi)
\hspace{10pt} ,
\end{equation}
where the $\vec k$ sum runs over the discrete spectrum $\vec k_{\vec \alpha}$
of the 3-torus.
Since $\psi_{-\vec k}(\vec x\,) = \psi_{\vec k}^\star(\vec x\,)$,
it is clear that the expansion coefficients $\Phi_{\vec k}(\eta)$ have to satisfy
$\Phi_{-\vec k}(\eta) = \Phi_{\vec k}^\star(\eta)$ so that a real representation of
the metric perturbation $\Phi(\eta,\vec x\,)$ is obtained.

The expansion coefficients $\Phi_{\vec k}(\eta)$ can be written as a product of
two factors.
The first factor is a Gaussian random number with zero mean and unit variance,
while the second describes the time-evolution in $\eta$,
which in turn is evolved from an initial condition at $\eta=0$
specified by the power spectrum ${\cal P}(k)$.

As it is done for $\Phi(\eta,\vec x\,)$,
the spatial covariant divergence of the velocity field of
the tightly coupled baryon-radiation component can by expanded as
\begin{equation}\label{velocity_divergence}
  V(\eta,\vec x\,) \; = \;
  \sum_{\vec k} \, V_{\vec k}(\eta)  \,
  \frac{\partial}{\partial r} \, \psi_{\vec k}(r,\theta,\phi)
\hspace{10pt} .
\end{equation}
Denoting the expansion coefficient of the relative perturbation in the
radiation component by $\delta_{\vec k}(\eta)$,
the Sachs-Wolfe formula \cite{Sachs_Wolfe_1967} can be written
in the tight-coupling approximation as
\begin{eqnarray}
   \hspace*{-40pt} \delta T(\hat n) & = & \nonumber
   \sum_{\vec k}\left[ \left( \Phi_{\vec k}(\eta) + \frac{\delta_{\vec k}(\eta)}4
    + \frac{a(\eta) V_{\vec k}(\eta)}{E_{\vec k}}\,
    \frac{\partial}{\partial r}\right) \, \psi_{\vec k}(r(\eta),\theta,\phi)
    \right]_{\eta=\eta_{\hbox{\scriptsize SLS}}}
  \\ & & \; \; + \; \label{SWE_expansion}
  2  \sum_{\vec k} \int_{\eta_{\hbox{\scriptsize SLS}}}^{\eta_0} d\eta
  \frac{\partial\Phi_{\vec k}(\eta)}{\partial\eta} \psi_{\vec k}(r(\eta),\theta,\phi)
\hspace{10pt} ,
\end{eqnarray}
where SLS denotes the surface of last scattering.
$\eta_{\hbox{\scriptsize{SLS}}}$ is the time of last scattering assumed in
the $\Lambda$CDM model to be at redshift $z\simeq 1089$,
and $\eta_0$ is the present time.

Requiring isentropic initial conditions,
the first two terms of the integrand in (\ref{SWE_expansion}) can be
approximated for modes outside the horizon at $\eta_{\hbox{\scriptsize{SLS}}}$ by
\begin{equation}\label{approximation_radiation}
  \Phi(\eta_{\hbox{\scriptsize{SLS}}},\vec x(\eta_{\hbox{\scriptsize{SLS}}})) \; + \;
  \frac 14 \, \delta(\eta_{\hbox{\scriptsize{SLS}}},\vec x(\eta_{\hbox{\scriptsize{SLS}}}))
  \; \simeq \;
  \frac 13 \, \Phi(\eta_{\hbox{\scriptsize{SLS}}},\vec x(\eta_{\hbox{\scriptsize{SLS}}}))
\hspace{10pt} .
\end{equation}
For large angular scales, one thus gets
\begin{eqnarray}
   \hspace*{-40pt} \delta T(\hat n) & = & \nonumber
   \sum_{\vec k}\left[ \left( \frac 13\,\Phi_{\vec k}(\eta)
    + \frac{a(\eta) V_{\vec k}(\eta)}{E_{\vec k}}\,
    \frac{\partial}{\partial r}\right) \, \psi_{\vec k}(r(\eta),\theta,\phi)
    \right]_{\eta=\eta_{\hbox{\scriptsize SLS}}}
  \\ & & \hspace{15pt} + \; \label{SWE_approx}
  2  \sum_{\vec k} \int_{\eta_{\hbox{\scriptsize SLS}}}^{\eta_0} d\eta
  \frac{\partial\Phi_{\vec k}(\eta)}{\partial\eta} \psi_{\vec k}(r(\eta),\theta,\phi)
\hspace{10pt} .
\end{eqnarray}
In order to understand the anisotropic behaviour of the variance $\sigma^2(\hat n)$,
we now ignore the integrated Sachs-Wolfe contribution,
the last term in (\ref{SWE_approx}).
This contribution cannot be the cause of the anisotropy
since it describes the evolution of the perturbations along
the line of sight from the surface of last scattering to the observer and
the points along the path are not related by the topology in general.

Let us now check whether already a simple argument based on the first term in
eq.(\ref{SWE_approx}) can describe the anisotropic behaviour.
Presuming that the sum over $\vec k$ runs only over one half
of the $\vec k$ space, the two terms belonging to $\vec k$ and
$-\vec k$ can be summarised with $a_{-\vec k}(\eta)=a_{\vec k}^\star(\eta)$ as
\begin{equation}
  \label{Phi}
  \Phi(\eta,\vec x\,) \; = \;
  \sum_{\vec k}\left(
            a_{\vec k}(\eta) \, e^{i \vec k \cdot \vec x}
            \, + \, a_{-\vec k}(\eta)  e^{-i \vec k \cdot \vec x}
            \right)
            \; = \;
            2\, \sum_{\vec k} \hbox{ Re}\left(
            a_{\vec k}(\eta) \, e^{i \vec k \cdot \vec x}\,\right)
\hspace{10pt} ,
\end{equation}
where the normalisation $N$ is absorbed by the coefficients $a_{\vec k}(\eta)$.
At the recombination time $\eta_{\hbox{\scriptsize{SLS}}}$,
the square of the temperature perturbation $ \delta T^2(\hat n)$
reads by taking only the first term in eq.(\ref{SWE_approx}) into account
\begin{eqnarray} \nonumber
  \hspace*{-40pt}\delta T^2(\hat n) & \simeq &
  \frac 19 \, \Phi^2(\eta_{\hbox{\scriptsize{SLS}}},\vec x\,)
  \; = \;
  \frac 49 \, \left(
  \sum_{\vec k} \hbox{Re}\left[
    a_{\vec k}(\eta_{\hbox{\scriptsize{SLS}}}) \, e^{i \vec k \cdot \vec x(\eta_{\hbox{\scriptsize{SLS}}})}\right]\;
    \right)^2
  \\ \nonumber & \hspace*{-20pt} = &\hspace{-10pt}
  \frac 49 \, \left(
  \sum_{\vec k} \hbox{Re}\left[a_{\vec k}(\eta_{\hbox{\scriptsize{SLS}}})\right] \, \cos(\vec k \cdot \vec x(\eta_{\hbox{\scriptsize{SLS}}})) - \hbox{Im}\left[a_{\vec k}(\eta_{\hbox{\scriptsize{SLS}}})\right] \, \sin(\vec k \cdot \vec x(\eta_{\hbox{\scriptsize{SLS}}}))
    \right)^2
    \\ \nonumber &  \hspace*{-20pt}\simeq &\hspace{-10pt}
    \frac 49 \,
    \sum_{\vec k} \Big(
    \hbox{Re}\left[a_{\vec k}(\eta_{\hbox{\scriptsize{SLS}}})\right]^2 \, \cos^2(\vec k \cdot \vec x(\eta_{\hbox{\scriptsize{SLS}}})) + \hbox{Im}\left[a_{\vec k}(\eta_{\hbox{\scriptsize{SLS}}})\right]^2 \, \sin^2(\vec k \cdot \vec x(\eta_{\hbox{\scriptsize{SLS}}})) 
   \\ & & \hspace{-10pt} \label{deltaT_NSW}
   - \,2\,\hbox{Re}\left[a_{\vec k}(\eta_{\hbox{\scriptsize{SLS}}})\right]\,
   \hbox{Im}\left[a_{\vec k}(\eta_{\hbox{\scriptsize{SLS}}})\right]\,
   \sin(\vec k \cdot \vec x(\eta_{\hbox{\scriptsize{SLS}}}))\,
   \cos(\vec k \cdot \vec x(\eta_{\hbox{\scriptsize{SLS}}}))
\; \Big) \hspace{5pt} .
\end{eqnarray}
In the last equation, only the diagonal term is retained.
Assuming $\left< \hbox{Re}(a_{\vec k})\,\hbox{Im}(a_{\vec k}) \right>=0$ and
$\left<\hbox{Re}(a_{\vec k})^2\right> = \left<\hbox{Im}(a_{\vec k})^2\right> := \Sigma(k)^2$ ,
the expectation value $\sigma^2(\hat n) = \left< \delta T^2(\hat n) \right>$,
eq.\,(\ref{variance_appr}), is obtained from the
ensemble average
\begin{equation} \label{expactation_T2}
  \left< \delta T^2(\hat n) \right> \; = \;
  \frac 49 \, 
  \sum_{\vec k} \Sigma(k)^2 \,
  \Big(\underbrace{\cos^2(\vec k \cdot \vec x(\eta_{\hbox{\scriptsize{SLS}}}))
    + \sin^2(\vec k \cdot \vec x(\eta_{\hbox{\scriptsize{SLS}}}))}_{=1} \; \Big)
\hspace{10pt} .
\end{equation}
So this approximation for an estimation of the Sachs-Wolfe formula leads to
an isotropic result.
Therefore this approximation is too crude and a more sophisticated
and refined approach is needed.


\section{The orbit sum rules}
\label{orbit-sum-rule}


In this section, we make use of orbit sum rules.
These relate a sum over eigenfunctions of the considered space
(see section \ref{torus_topology}) with a sum over orbits,
whereas the orbits are geodesic paths between two given points in the
considered space.
They find applications in the study of eigenfunctions,
for example in the field of quantum chaos
\cite{Aurich_Steiner_1991,Aurich_Steiner_1995,Aurich_Steiner_2001}.
One prominent example is the question for enhanced amplitudes of
the eigenfunctions in special regions due to classical periodic orbits:
the scar phenomenon \cite{Heller_1984}.

\begin{figure}
\begin{center}
\vspace*{0pt}
\begin{minipage}{15.8cm}
\hspace*{0pt}\includegraphics[width=15.8cm]{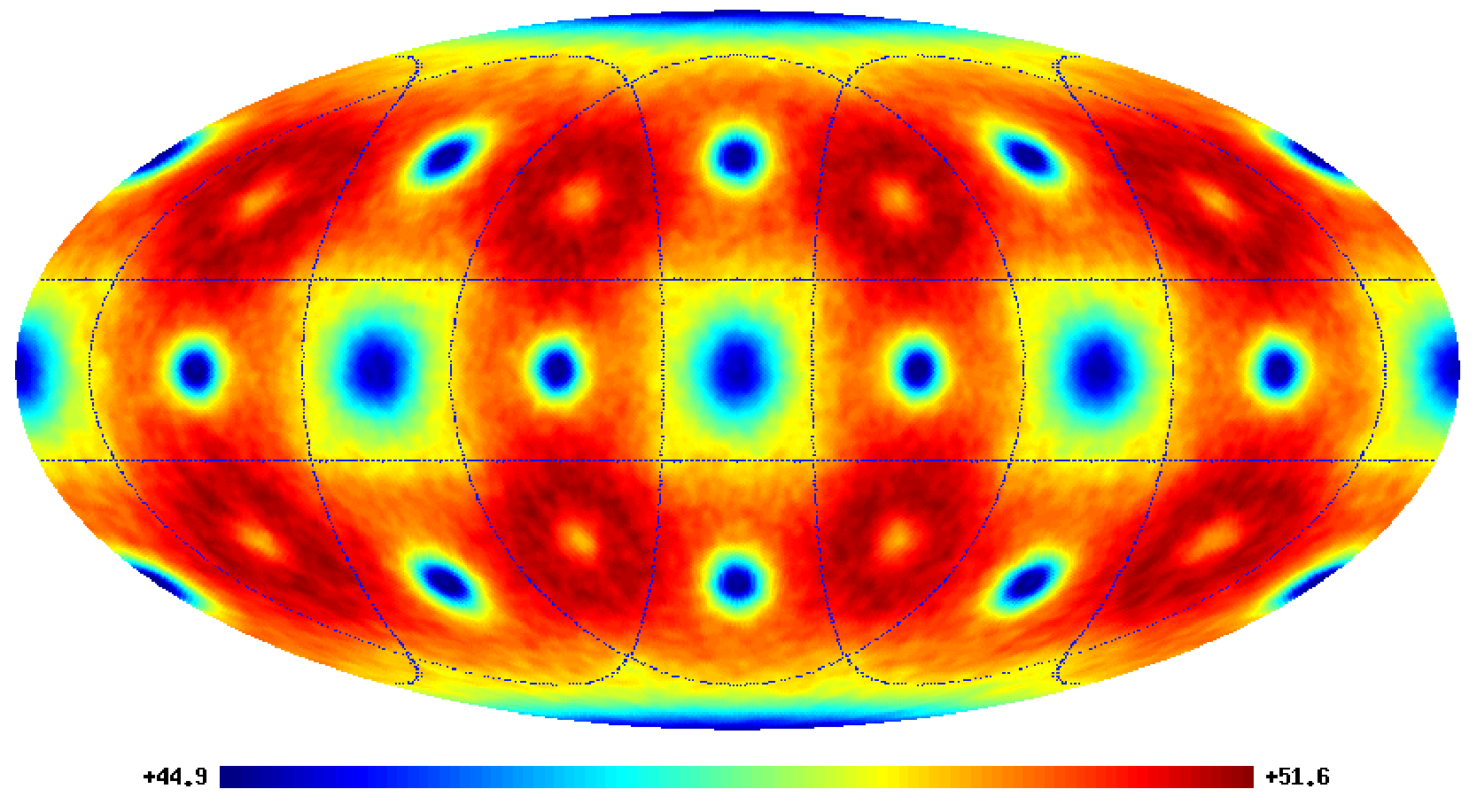}
\end{minipage}
\end{center}
\vspace*{0pt}
\caption{\label{Fig:L2_Generatoren}
  The standard deviation $\sigma(\hat n)$ already shown in
  figure \ref{Fig:L2_StdDev} is presented in addition to the
  circles-in-the-sky corresponding to the generators $\gamma$ of
  the cubic 3-torus group $\Gamma$ with $L=2$.
  This reveals the faces of the directly neighbouring torus cells.
  The sky map is smoothed with a Gaussian of $\hbox{fwhm} = 2^\circ$.
}
\end{figure}

\begin{figure}
\begin{center}
\vspace*{0pt}
\begin{minipage}{15.8cm}
\hspace*{0pt}\includegraphics[width=15.8cm]{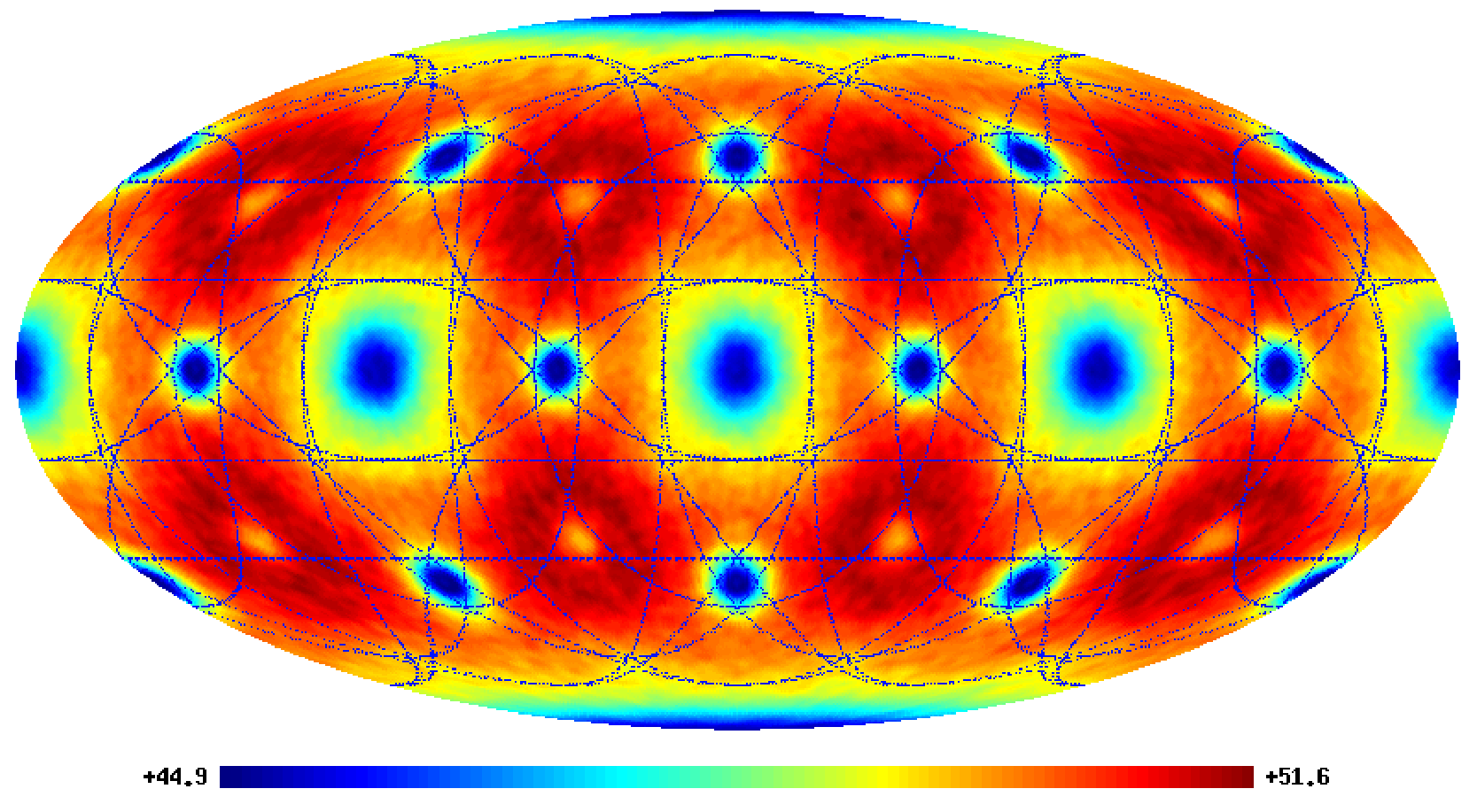}
\end{minipage}
\end{center}
\vspace*{0pt}
\caption{\label{Fig:L_2_Zweierprodukte}
The same as figure \ref{Fig:L2_Generatoren},
but now also the circles-in-the-sky corresponding to the products
of two generators are included.
This reveals not only the boundary lines of directly neighbouring
topological cells, but also the next ones.
}
\end{figure}

That the periodic patterns appearing in figures
\ref{Fig:L2_StdDev} and \ref{Fig:L2_StdDev_withDip},
illustrating the violation of isotropy 
are related to the orientation of the fundamental cell
becomes obvious,
when the boundary lines of the fundamental cell are plotted
together with the standard deviation $\sigma(\hat n)$.
To this aim, figure \ref{Fig:L2_Generatoren} displays in addition to
$\sigma(\hat n)$ the circles-in-the-sky \cite{Cornish_Spergel_Starkman_1998b}
corresponding to the generators $\gamma$ of the cubic 3-torus group $\Gamma$.
This shows the intersections of the surface of last scattering with the planes, 
which are defined by contact faces between the fundamental cell and its directly
neighbouring torus cells.
The figure \ref{Fig:L_2_Zweierprodukte} also shows the circles-in-the-sky
due to the product of two generators.
This clearly demonstrates the relation of the $\sigma(\hat n)$ anisotropy
structures with the cubic 3-torus group $\Gamma$.

The anisotropy in $\sigma(\hat n)$ should be explained by the
Sachs-Wolfe formula.
As discussed in the previous section, 
the metric perturbation $\Phi(\eta,\vec x\,)$ is expanded
with respect to the eigenfunctions (\ref{eigenfunction})
and that is the point where one can make use of the orbit sum rules.
The orbit sum rule for the 3-torus reads \cite{Steiner_2016}
\begin{equation}\label{OSR}
  \sum_{\vec \alpha} h(k_{\vec \alpha}) \, \psi_{\vec \alpha}(\vec x\,) \,
  \psi^\star_{\vec \alpha}(\vec x\,')
  \; = \;
  \frac 1{2\pi^2} \, \sum_\gamma \frac{\widetilde{h}(d(\vec x,\gamma(\vec x\,'))}
        {d(\vec x,\gamma(\vec x\,'))}
  \hspace{10pt} ,
\end{equation}
with
\begin{equation}\label{OSR_hy}
\widetilde{h}(y) \; := \; \int_0^\infty \, k \, \sin(k\,y) \, h(k) \, dk
\hspace{10pt} ,
\end{equation}
the derivation of which is given in the Appendix.
Here, $d(\vec x,\gamma(\vec x\,'))$ denotes the Euclidean distance of the point
$\vec x$ from the mirror point $\gamma(\vec x\,')$ of $\vec x\,'$
due to the group element $\gamma\in\Gamma$.
The sum on the left-hand-side over the eigenmodes excludes
the zero mode $\vec\alpha=(0,0,0)$, and analogously,
the sum on the right-hand-side over the group elements $\gamma$
omits the identity.
These two contributions correspond to the monopole in the CMB or in
the anisotropy in $\sigma(\hat n)$.

In order to ensure the absolute convergence of the sums,
the following conditions are required (see the Appendix):
\begin{itemize}
\item[(i)] $h(k) \; = \; h(-k)$
\item[(ii)] $h(k)$ is holomorphic around a strip of the positive $k$ axis
  with $\epsilon \leq k \leq\infty$, $\epsilon>0$
\item[(iii)] $h(k) = O(1/(|k|^{3+\delta}))$ for $|k|\to \infty$,
  $\delta>0$ \hspace{10pt} .
\end{itemize}
If these conditions are not satisfied, one can in some cases regularise
the sums, see below the contribution to the Doppler case.

From (\ref{OSR}) follows for $\vec x\,'=0$ the special case
\begin{equation}\label{OSR_one_psi}
  \sum_{\vec \alpha} h(k_{\vec \alpha}) \, \psi_{\vec \alpha}(\vec x\,)
  \; = \;
  \frac{\sqrt V}{2\pi^2} \, \sum_\gamma \frac{\widetilde{h}(d(\vec x,\gamma(\vec 0\,))}
        {d(\vec x,\gamma(\vec 0\,))}
  \hspace{10pt} ,
\end{equation}
since $\psi_{\vec \alpha}(\vec x\,'=0) = 1/\sqrt V = L^{-3/2}$.
The special case $\vec x\,'=0$ is important since this point is
identical to the position of the CMB observer.

Ideally one would set the function $h(k)$ in eq.(\ref{OSR_one_psi})
equal to the expansion coefficients $\Phi_{\vec k}(\eta)$
of the metric perturbation  $\Phi(\eta,\vec x\,)$,
see eq.(\ref{metric_perturbation}).
However, the function $\widetilde{h}(y)$ should be known analytically
and because $\Phi_{\vec k}(\eta)$ contains the complicated CMB physics,
we restrict ourselves here to a simple case.
Because small scale structures in $\delta T(\hat n)$ are suppressed by the
Gaussian smoothing in our analysis,
this suppression carries over to a Gaussian suppression of eigenfunctions
$\psi_{\vec k}(\vec x\,)$ belonging to large wave numbers $|\vec k\,|$,
and thus, the Gaussian smoothing is taken into account by the function
\begin{equation}\label{fkt_heat_kernel}
  h(k) \; = \; e^{-k^2/(2\sigma^2)}
  \hspace{10pt} .
\end{equation}
The integral (\ref{OSR_hy}) for $\widetilde{h}(y)$ with
(\ref{fkt_heat_kernel}) evaluates to
\begin{equation}\label{fkt_heat_kernel_tilde}
\widetilde{h}(y) \; = \;
\sqrt{\frac\pi 2}\; \sigma^3 \; y \; \e^{-\frac{\sigma^2\,y^2}{2}}
\hspace{10pt} .
\end{equation}
This does ignore the power spectrum  ${\cal P}(k)$ and the fact
that the modes are weighted by Gaussian random numbers as discussed
in section \ref{Sachs-Wolfe},
but if this $h(k)$ already leads to structures analogous to
figure \ref{Fig:L2_StdDev},
the correct expansion coefficients $\Phi_{\vec k}(\eta)$ will inherit
these structures.

Using the function pair (\ref{fkt_heat_kernel}) and
(\ref{fkt_heat_kernel_tilde}) in (\ref{OSR}) leads to
\begin{equation}\label{OSR_heat_kernel}
  \sum_{\vec\alpha} e^{-\frac{k_{\vec \alpha}^2}{2\sigma^2}} \,
  \psi_{\vec \alpha}(\vec x\,) \, \psi^\star_{\vec \alpha}(\vec x\,')
  \; = \;
  \left(\frac{\sigma^2}{2\pi}\right)^{3/2} \,
  \sum_\gamma \exp\left(-\frac{\sigma^2}2\, d^2(\vec x,\gamma(\vec x\,'))\right)
  \hspace{10pt} .
\end{equation}
Note that this relation with $t := 1/(2\sigma^2) > 0$ is identical to
the heat kernel $\Theta_\Gamma(\vec x, \vec x\,'; t)$ of
a general 3-torus $\mathbb{T}^3$ satisfying the heat equation 
$\frac{\partial}{\partial t} \Theta_\Gamma = \Delta_{\vec x} \Theta_\Gamma$.
The special case (\ref{OSR_one_psi}) results in
\begin{equation}\label{OSR_heat_kernel_1p}
  \sum_{\vec\alpha} e^{-\frac{k_{\vec \alpha}^2}{2\sigma^2}} \,
  \psi_{\vec \alpha}(\vec x\,) 
  \; = \;
  \sqrt{V}\,\left(\frac{\sigma^2}{2\pi}\right)^{3/2} \,
  \sum_\gamma \exp\left(-\frac{\sigma^2}2\, d^2(\vec x,\gamma(\vec 0\,))\right)
  \hspace{10pt} .
\end{equation}

Ignoring the integrated Sachs-Wolfe contribution in (\ref{SWE_approx})
as discussed in section \ref{Sachs-Wolfe},
one gets the large angle approximation
\begin{equation}
   \hspace*{-40pt} {\cal G}(\delta T(\hat n) ) \; = \;
   \sum_{\vec k}\left[ \left( \frac 13\,\Phi_{\vec k}(\eta)
    + \frac{a(\eta) V_{\vec k}(\eta)}{E_{\vec k}}\,
    \frac{\partial}{\partial r}\right) \,  e^{-\frac{k_{\vec \alpha}^2}{2\sigma^2}}\,
    \psi_{\vec k}(r(\eta),\theta,\phi)
    \right]_{\eta=\eta_{\hbox{\scriptsize SLS}}}
\hspace{3pt} ,
\end{equation}
where ${\cal G}(\delta T(\hat n))$ denotes the Gaussian smearing of the
temperature fluctuations $\delta T(\hat n)$.
As announced, we set $\Phi_{\vec k}(\eta) = A$ and
$a(\eta) V_{\vec k}(\eta) =B$ independent of $\vec k$.
Then one obtains
\begin{eqnarray}
   \hspace*{-40pt} {\cal G}(\delta T(\hat n) ) & = & \nonumber
   \sum_{\vec k}\left[ \left( \frac A3\,
    + \frac{B}{k^2}\,
    \frac{\partial}{\partial\tau}\right) \, e^{-\frac{\vec k^2}{2\sigma^2}}\,
    \psi_{\vec k}(\tau(\eta),\theta,\phi)
    \right]_{\eta=\eta_{\hbox{\scriptsize SLS}}}
   \\ & = & \nonumber
   \frac A3\,\sum_{\vec k} e^{-\frac{\vec k^2}{2\sigma^2}}
   \psi_{\vec k}(\vec x(\eta_{\hbox{\scriptsize{SLS}}})) \, + \,
   B \, \frac{\partial}{\partial r} \,
   \sum_{\vec k} \frac 1{\vec k^2} \, e^{-\frac{\vec k^2}{2\sigma^2}}
   \psi_{\vec k}(\vec x(\eta_{\hbox{\scriptsize{SLS}}}))
   \\ & = & \nonumber
   \frac A3\, \sqrt{V}\,\left(\frac{\sigma^2}{2\pi}\right)^{3/2} \,
   \sum_\gamma \exp\left(-\frac{\sigma^2}2\,
   d^2(\vec x(\eta_{\hbox{\scriptsize{SLS}}}),\gamma(\vec 0\,))\right)
   \\ & & \nonumber \; \; + \,
   B \, \frac{\partial}{\partial r} \,
   \sum_{\vec k} \frac 1{\vec k^2} \, e^{-\frac{\vec k^2}{2\sigma^2}}
   \psi_{\vec k}(\vec x(\eta_{\hbox{\scriptsize{SLS}}}))
   \hspace{10pt} .
\end{eqnarray}
While the term proportional to $A$ is analysed by $h(p)$
as defined in (\ref{fkt_heat_kernel}),
one needs for the Doppler term proportional to $B$ the function
\begin{equation}\label{fkt2_heat_kernel}
  h(k) \; = \; \frac 1{k^2}e^{-k^2/(2\sigma^2)}
  \hspace{10pt} .
\end{equation}
In this case, the corresponding function $\widetilde{h}(y)$ has to be
regularised by replacing $\widetilde{h}(y)$ by
$\widetilde{h}_{\hbox{\scriptsize{reg}}}(y)$
which has no contribution due to the zero mode (see the Appendix)
and thus $(y>0)$
\begin{equation}\label{fkt2_heat_kernel_tilde}
  \widetilde{h}_{\hbox{\scriptsize{reg}}}(y) \; = \;
  -\;\frac{\pi}2\; \hbox{erfc}\left(\frac{\sigma}{\sqrt{2}}\,y\right)
  \hspace{10pt} .
\end{equation}

Let us now write
${\cal G}(\delta T(\hat n)) = {\cal G}(\delta T^{\hbox{\scriptsize SW}}(\hat n)) +
{\cal G}(\delta T^{\hbox{\scriptsize Dop}}(\hat n))$,
where the first summand denotes the contribution due to the
perturbation in the metric
\begin{equation}\label{Fluctuation_SW}
  {\cal G}(\delta T^{\hbox{\scriptsize SW}}(\hat n)) \; = \;
  \frac A3\, \sqrt{V}\,\left(\frac{\sigma^2}{2\pi}\right)^{3/2} \,
  \sum_\gamma \exp\left(-\frac{\sigma^2}2\, d^2(\vec x(\eta_{\hbox{\scriptsize{SLS}}}),\gamma(\vec 0\,))\right)
\end{equation}
and the second one is due to the Doppler effect
\begin{equation}
{\cal G}(\delta T^{\hbox{\scriptsize Dop}}(\hat n)) \; = \;
   -\, \frac{\sqrt{V}}{2\pi^2}\, B\, \frac{\pi}2\,
            \frac{\vec x(\eta_{\hbox{\scriptsize{SLS}}})}
            {|\vec x(\eta_{\hbox{\scriptsize{SLS}}})|}
   \cdot \hbox{ grad}_{\vec x}
   \left. \sum_\gamma 
   \frac{\hbox{erfc}\left(\frac{\sigma}{\sqrt{2}}\,d(\vec x,\gamma(\vec 0\,)\right)}
        {d(\vec x,\gamma(\vec 0\,)}
        \right|_{\vec x = \vec x(\eta_{\hbox{\scriptsize{SLS}}})}
\end{equation}
The contribution ${\cal G}(\delta T^{\hbox{\scriptsize Dop}}(\hat n))$ can be
evaluated with
$$
     \frac{\partial}{\partial y}\,\left(
     \frac{\hbox{erfc}\Big(\frac\sigma{\sqrt 2}\, y\Big)}{y}\right)
     \; = \;
     -\, \sqrt{\frac 2\pi}\, \frac\sigma y \, e^{-\frac 12 \sigma^2 y^2} \, - \,
     \frac 1{y^2}\hbox{erfc}\Big(\frac\sigma{\sqrt 2}\, y\Big)
$$
     which leads with the abbreviation
     $\tau_\gamma := d(\vec x(\eta_{\hbox{\scriptsize{SLS}}}),\gamma(\vec 0\,))$ to
\begin{eqnarray}
   \hspace*{-40pt} {\cal G}(\delta T^{\hbox{\scriptsize Dop}}(\hat n) ) & = & \nonumber
   \frac{\sqrt{V}}{4\pi}\, B\,
   \sum_\gamma \left(
     \sqrt{\frac 2\pi}\, \frac\sigma\tau_\gamma \, e^{-\frac 12 \sigma^2 \tau_\gamma^2} \, + \,
     \frac 1{\tau_\gamma^2}\hbox{erfc}\Big(\frac\sigma{\sqrt 2}\, \tau_\gamma\Big)
     \right)
    \\ & & \label{Fluctuation_Dop} \hspace*{100pt} \; \; \times
    \underbrace{\frac{\vec x(\eta_{\hbox{\scriptsize{SLS}}})}
            {|\vec x(\eta_{\hbox{\scriptsize{SLS}}})|}}_{= \hat n}
   \cdot \hbox{ grad}_{\vec x} \, d(\vec x(\eta_{\hbox{\scriptsize{SLS}}}),\gamma(\vec 0\,))
\hspace{10pt} .
\end{eqnarray}
For the cubic 3-torus with side-length $L$,
the transformations $\gamma(\vec 0\,)$ are given by
$\gamma(\vec 0\,) \; = \; \vec \beta_\gamma \, L$ with
$\vec\beta_\gamma=(\beta_1,\beta_2,\beta_3) \in \mathbb{Z}^3$ and thus
\begin{eqnarray} \nonumber
\tau_\gamma \; = \; d(\vec x,\gamma(\vec 0\,)) & = &
|\vec x - \vec \beta_\gamma \, L|
\\ & = & \nonumber
\sqrt{(x_1-\beta_1 L)^2 +(x_2-\beta_2 L)^2 +(x_3-\beta_3 L)^2}
\hspace{10pt} .
\end{eqnarray}
The directional derivative results simply in
$$
\vec x \cdot \hbox{ grad}_{\vec x} \; d(\vec x,\gamma(\vec 0\,)) \; = \;
\frac{x_1(x_1-\beta_1 L) +  x_2(x_2-\beta_2 L) + x_3(x_3-\beta_3 L)}
{\sqrt{(x_1-\beta_1 L)^2 +(x_2-\beta_2 L)^2 +(x_3-\beta_3 L)^2}}
\hspace{10pt} .
$$
This allows the evaluation of the approximate temperature
fluctuation of the Sachs-Wolfe contribution (\ref{Fluctuation_SW})
due to the metric perturbation
and the Doppler contribution (\ref{Fluctuation_Dop}).

The geometrical interpretation follows from the crucial quantity
$\tau_\gamma \; = \; d(\vec x(\eta_{\hbox{\scriptsize{SLS}}}),\gamma(\vec 0\,))$
in the equations (\ref{Fluctuation_SW}) and (\ref{Fluctuation_Dop}),
which is the distance of the point $\vec x$ on the surface of last scattering
to the mirror points $\gamma(\vec 0\,)$ of the observer position
due to the topology.
With increasing size $L$ of the 3-torus cell,
these distances also increase until the transformed points are
significantly beyond the surface of last scattering.
In this case, the anisotropy produced by the group $\Gamma$ vanishes
and only the neutral group element yields a contribution,
i.\,\,e.\ the distance to the original observer position which is nothing
else than the radius of the surface of last scattering.
For 3-torus cells small enough,
such that the surface of last scattering does not lie completely within
the torus cell,
the relevant group elements reveal the topological signature.

\begin{figure}
\begin{center}
\vspace*{0pt}
\begin{minipage}{8cm}
\hspace*{-50pt}\includegraphics[width=6.0cm,angle=270]{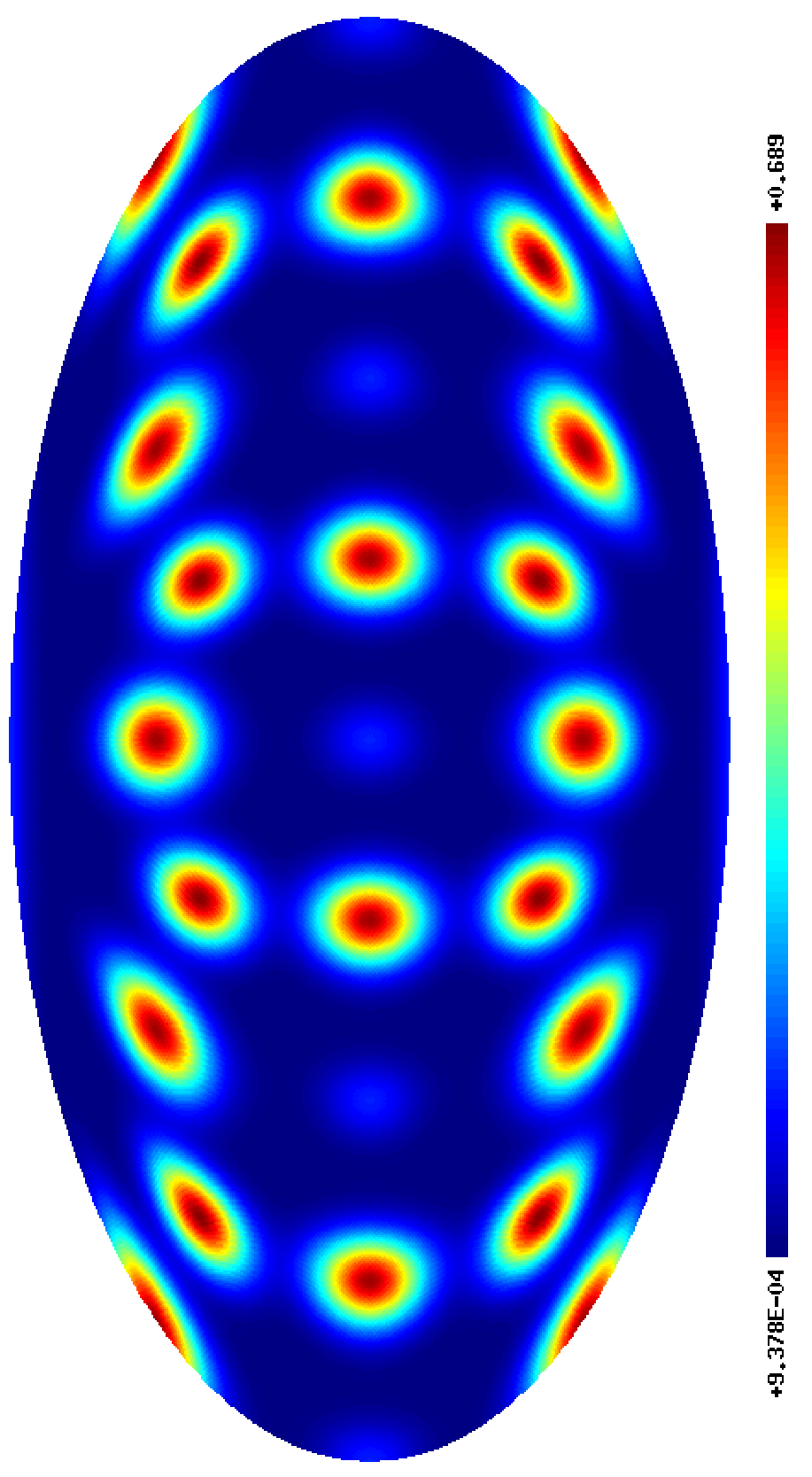}
\end{minipage}
\end{center}
\vspace*{0pt}
\caption{\label{Fig:SW_contribution}
  The contribution ${\cal G}(\delta T^{\hbox{\scriptsize SW}}(\hat n))$
  eq.(\ref{Fluctuation_SW}) is shown for the torus with $L=2$ for $\sigma=2.5$.
  Here, we set $A=1$.
}
\end{figure}

\begin{figure}
\begin{center}
\vspace*{0pt}
\begin{minipage}{8cm}
\hspace*{-50pt}\includegraphics[width=6.0cm,angle=270]{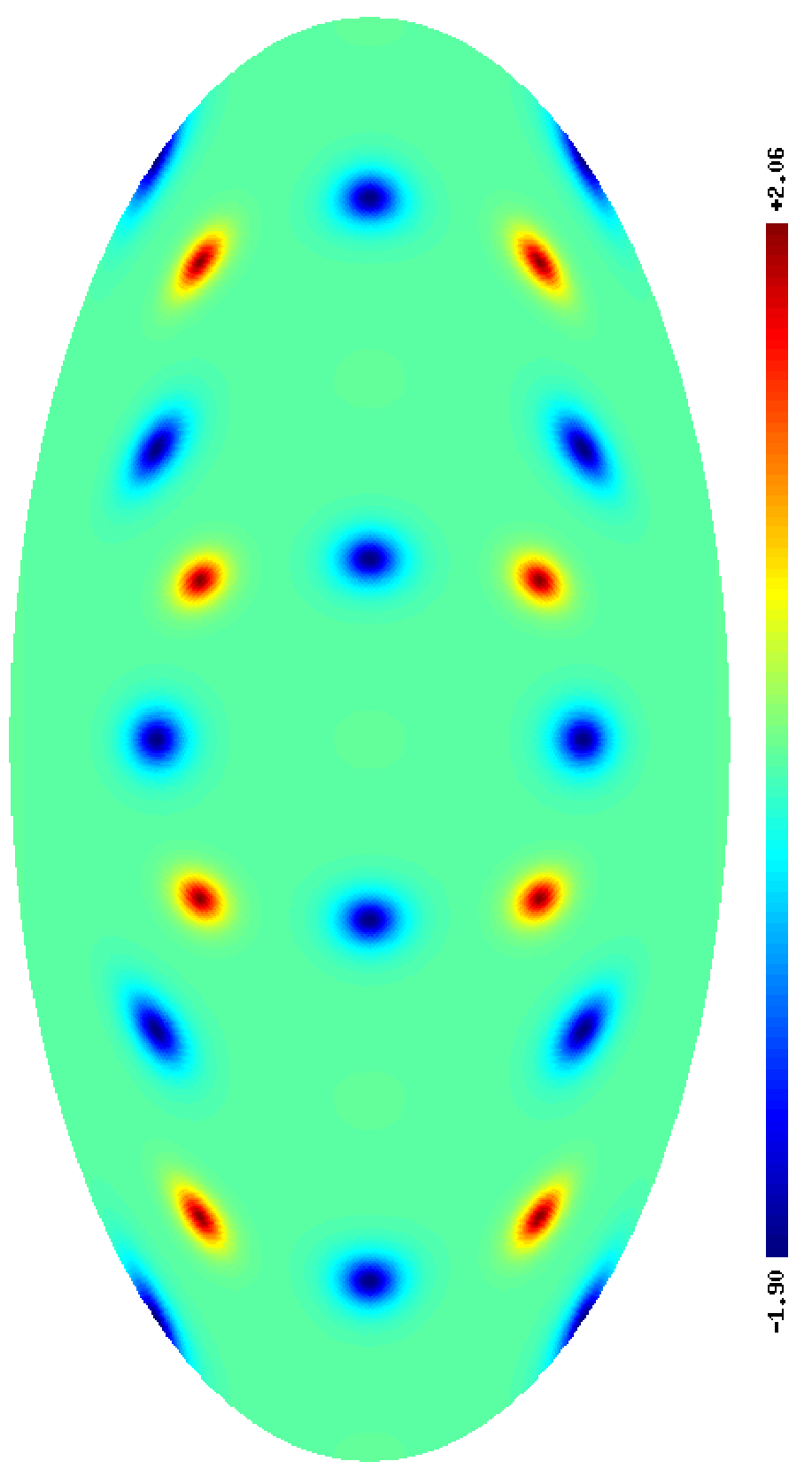}
\end{minipage}
\end{center}
\vspace*{0pt}
\caption{\label{Fig:Dop_contribution}
  The contribution ${\cal G}(\delta T^{\hbox{\scriptsize Dop}}(\hat n))$
  eq.(\ref{Fluctuation_Dop}) is shown for the torus with $L=2$ for $\sigma=2.5$.
  Here, we set $B=1$.
}
\end{figure}

Figure \ref{Fig:SW_contribution} shows the Sachs-Wolfe contribution
${\cal G}(\delta T^{\hbox{\scriptsize SW}}(\hat n))$ due to the metric perturbation,
eq.(\ref{Fluctuation_SW}), for the cubic 3-torus with side-length $L=2$,
where a smoothing of $\sigma=2.5$ is used.
The Doppler contribution ${\cal G}(\delta T^{\hbox{\scriptsize Dop}}(\hat n))$
given in eq.(\ref{Fluctuation_Dop}) is shown in figure
\ref{Fig:Dop_contribution} again for the torus with $L=2$,
and $\sigma=2.5$ is used as in figure \ref{Fig:SW_contribution}.
The dipole is not removed in both figures and thus,
one should compare the structures seen in figures \ref{Fig:SW_contribution}
and \ref{Fig:Dop_contribution} with the structures revealed in
figure \ref{Fig:L2_StdDev_withDip}, where the whole CMB physics is included.
One cannot expect that an appropriate sum of figures \ref{Fig:SW_contribution}
and \ref{Fig:Dop_contribution} can reproduce the behaviour of the
CMB simulations shown in figure \ref{Fig:L2_StdDev_withDip},
since the power spectrum ${\cal P}(k)$ is ignored.
However, this analysis demonstrates the origin of the violation of isotropy
in the case of the cubic 3-torus universes.

\section{The Sachs-Wolfe approximation with an almost
  Harrison-Zeldovitch spectrum}

\label{NSW-with-almost-Harrison-Zeldovitch}

\begin{figure}
\begin{center}
\vspace*{0pt}
\begin{minipage}{8cm}
\hspace*{-50pt}\includegraphics[width=6.0cm,angle=270]{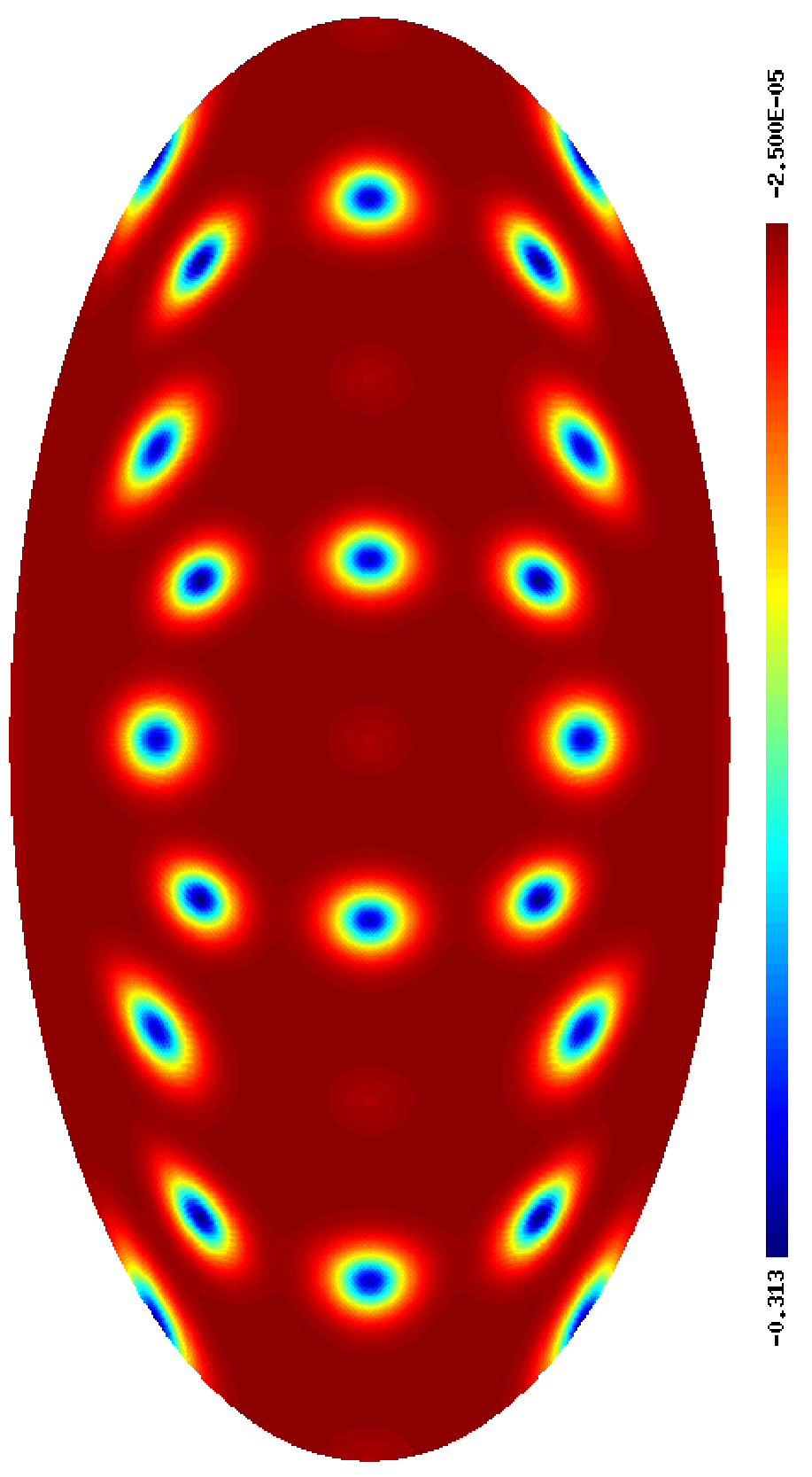}
\end{minipage}
\end{center}
\vspace*{0pt}
\caption{\label{Fig:sw_k2_contribution}
  The contribution ${\cal G}(\delta T^{\hbox{\scriptsize SW}}(\hat n))$
  computed from equation (\ref{Fluctuation_SW_km2}) is shown for
  the torus with $L=2$ for $\sigma=2.5$, we set $A=1$.
}
\end{figure}

The above evaluation ignores the power spectrum ${\cal P}(k)$,
which weights the modes on the left-hand-side of equation (\ref{OSR}).
The spectral index $n_s$ is slightly lower than one.
Taking as the spectral index $n_s=1$ leads to the famous
Harrison-Zeldovitch power spectrum and one is led to ${\cal P}(k)\sim k^{-3/2}$.
For this choice, one would need
$h(p) = \exp(-k^2/(2\sigma^2)) / k^3$ for the Sachs-Wolfe contribution
due to the metric perturbation and
$h(p) = \exp(-k^2/(2\sigma^2)) / k^5$ for the Doppler term.
Since this does not lead to an analytical function $\widetilde{h}(y)$,
we consider here at least as an improvement to the previous calculation
of the contribution due to the metric perturbation, the choice
$h(p) = \exp(-k^2/(2\sigma^2)) / k^2$,
which leads to the function pair (\ref{fkt2_heat_kernel}) and
(\ref{fkt2_heat_kernel_tilde}).
The orbit-sum-rule (\ref{OSR_one_psi}) reads then
\begin{equation}\label{Fluctuation_SW_km2}
  {\cal G}(\delta T^{\hbox{\scriptsize SW}}(\hat n)) \; = \;
  - \, \frac A3 \, \frac{\sqrt{V}}{4\pi} \, \sum_\gamma
  \frac{\hbox{erfc}\left(\frac{\sigma}{\sqrt 2}\,
       d(\vec x(\eta_{\hbox{\scriptsize{SLS}}}),  \gamma(\vec 0\,))\right)}
  {d(\vec x(\eta_{\hbox{\scriptsize{SLS}}}),  \gamma(\vec 0\,))d(\vec x(\eta_{\hbox{\scriptsize{SLS}}}),  \gamma(\vec 0\,))}
  \hspace{10pt} .
\end{equation}
The numerical evaluation of equation (\ref{Fluctuation_SW_km2}) is shown
in figure \ref{Fig:sw_k2_contribution}.
The comparison of figure \ref{Fig:sw_k2_contribution} with
figure \ref{Fig:SW_contribution} shows the influence of the power spectrum
and demonstrates the persistence of the topological signature.

\section{The normalized modulus $\rho$ of the gradient of the temperature field $\delta T(\hat n)$}

\label{gradient-temperature-field}


In \cite{Aurich_Buchert_France_Steiner_2021,Aurich_Steiner_2024} it is shown
that the normalised modulus of the gradient of the temperature field
$\delta T(\hat n)$ denoted by $\rho$ betrays the size of the
cubic 3-torus cell by a linear dependence $\rho(L)$ on the torus size $L$
over a wide range of $L\in[1,3]$.
In that analysis the orientation of the torus cell with respect to the mask,
which is enforced by a comparison with observations, is held fixed.
In order to test the stability of this linear relation,
the analysis presented in \cite{Aurich_France_Steiner_2026} put the focus on
the variation of $\rho$ with respect to the mask orientation.
It turns out that the corresponding variation in $\rho$ is tiny
compared to the changes due to the torus size $L$,
so that the results in \cite{Aurich_Buchert_France_Steiner_2021} are robust.
This points to an almost negligible anisotropy in the 3-torus simulations.

The analysis in \cite{Aurich_Buchert_France_Steiner_2021} assumes
that the Universe is isotropic on average allowing to consider
\begin{equation}\label{def_rho}
  \rho \; := \;
  \frac{
  \frac 1{4\pi}\int_{{\cal S}^2} d^2\hat n \; \sigma_1(\hat n)}{
  \frac 1{4\pi}\int_{{\cal S}^2} d^2\hat n \; \sigma(\hat n)}
  \hspace{10pt} ,
\end{equation}
and thus ignoring a possible violation of isotropy.
Since the previous sections show that this is not rigorously fulfilled,
we extend the analysis of \cite{Aurich_Buchert_France_Steiner_2021}
with this point of view in this section.
Therefore, this quantity is now considered with respect to its
direction dependency as in \cite{Aurich_France_Steiner_2026}, that is
\begin{equation}\label{def_rho_directional}
  \rho(\hat n) \; := \;
  \frac{\sigma_1(\hat n)}{\sigma(\hat n)}
  \hspace{10pt} ,
\end{equation}
where $\sigma(\hat n)$ is defined in (\ref{def_variance}) and
$\sigma_1(\hat n)$  in (\ref{def_sigma_1}).

The Planck mask {\it COM\_Mask\_CMB-union\_2048\_R1.10.fits} available
at \url{https://irsa.ipac.caltech.edu/data/Planck/} is used,
which is the union of the several confidence masks and accepts 73\% of the sky
and is denoted U73 \cite{Ade_et_al_2014}.
In our mask-sweeping algorithm the Planck mask U73 is rotated
before it is applied to the unrotated torus simulations.
The reverse procedure that is applying the unrotated mask to the rotated
torus sky maps, would unnecessarily increase the computational costs.
In this way, a different number of maxima or minima as shown
in figures \ref{Fig:L2_StdDev} and \ref{Fig:L2_StdDev_withDip} are masked
and lead therefore to slightly different values of $\rho$.
The rotation of the mask is carried out by the HealPix 
\cite{Gorski_Hivon_Banday_Wandelt_Hansen_Reinecke_Bartelmann_2005}
routine {\it rotate\_alm}
(see \url{https://healpix.sourceforge.io/html/sub_rotate_alm.htm})
using their definition of the rotation angles.
This first rotation $\psi_{\hbox{\scriptsize healpix}}$ is performed around the polar axis ($z$-axis),
while the second rotation $\theta_{\hbox{\scriptsize healpix}}$ is around the
original (unrotated) $y$-axis.
A third Euler rotation $\varphi_{\hbox{\scriptsize healpix}}$ is not taken
into account due to the already
huge computational effort for the analysis of two Euler angles.
But this mask-sweeping already reveals the variation of $\rho$ with respect to
the orientation of the torus cell relative to the mask.
To be in line with the publication \cite{Aurich_France_Steiner_2026},
we set $\psi_{\hbox{\scriptsize healpix}}\equiv\Phi_{\textrm{\scriptsize or}}$ and
$\theta_{\hbox{\scriptsize healpix}}\equiv\Theta_{\textrm{\scriptsize or}}$.
 
\begin{figure}
\begin{center}
\vspace*{0pt}
\hspace*{20pt}\includegraphics[scale=0.20]{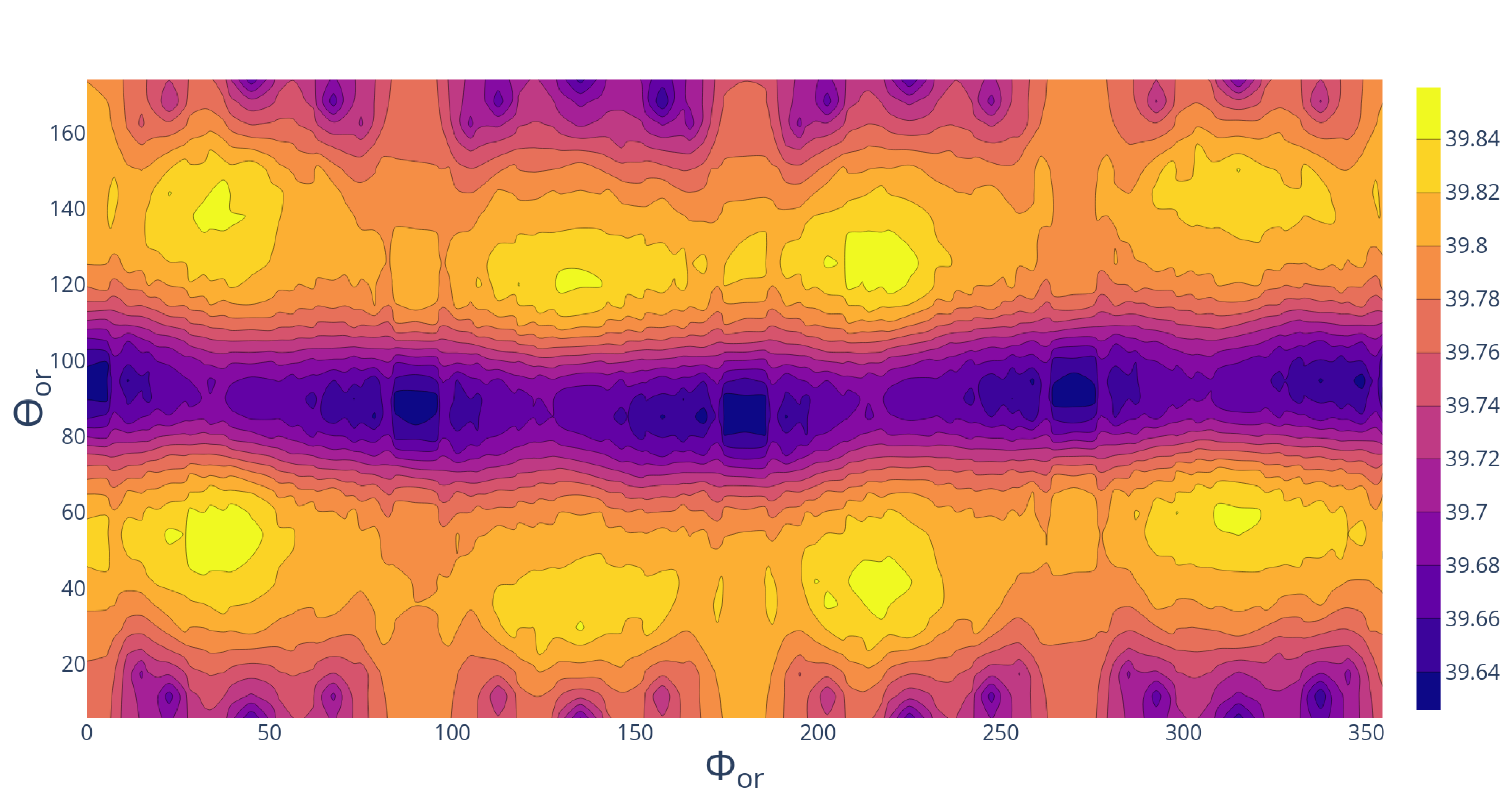}
\vspace*{-20pt}
\end{center}
\caption{\label{Fig:L2_carpet_Euler_angle}
  The variation of $\rho(\hat n)$ due to the CMB mask-sweeping method
  is shown for the torus ensemble with $L=2$
  in dependence of the Euler angles $\Phi_{\textrm{\scriptsize or}}$ and
  $\Theta_{\textrm{\scriptsize or}}$.
  One receives that the variation of the $\rho$ values is small.}
\end{figure}

\begin{figure}
\begin{center}
\hspace*{20pt}\includegraphics[scale=0.20]{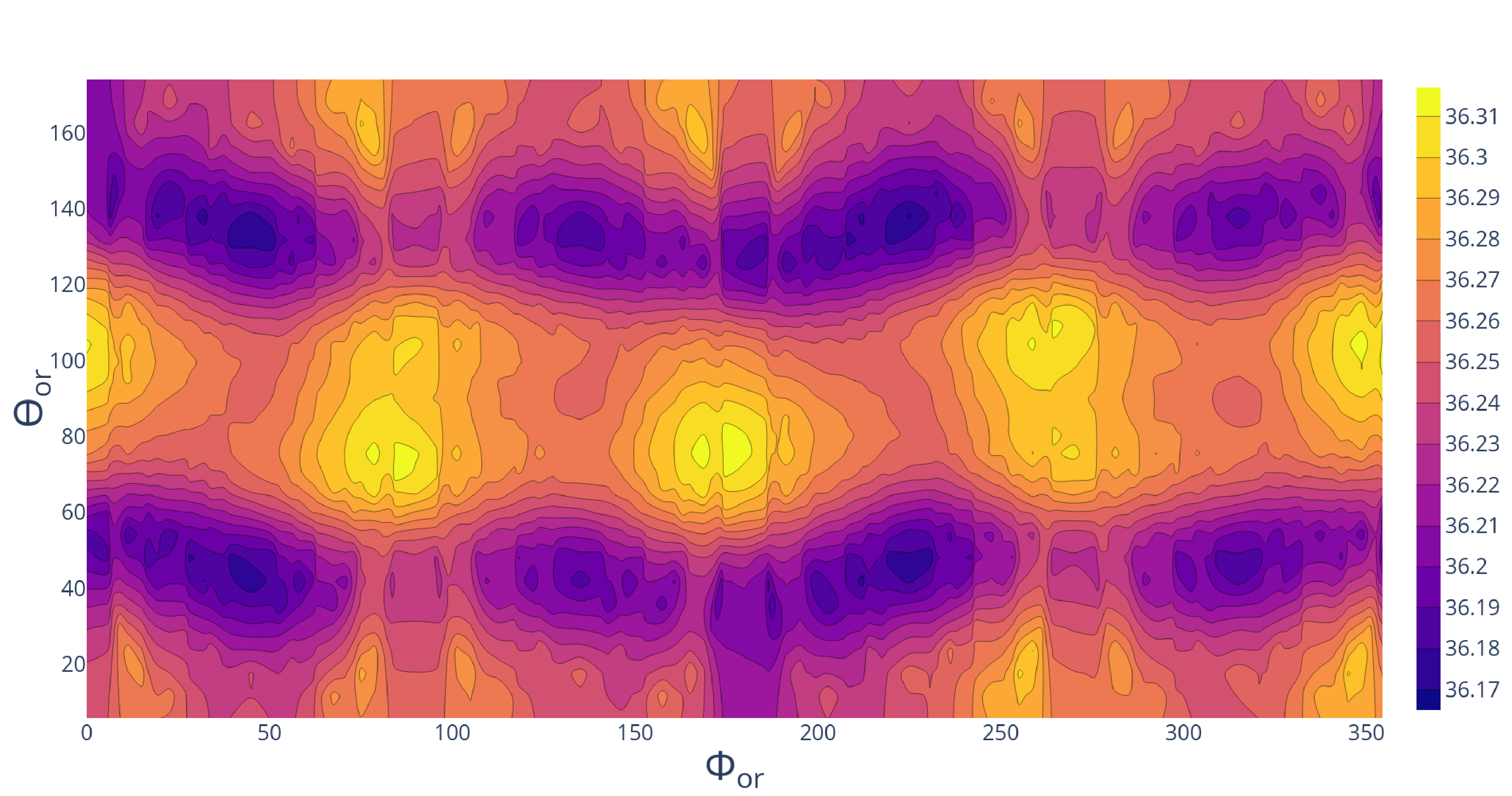}
\vspace*{-20pt}
\end{center}
\caption{\label{Fig:L3_carpet_Euler_angle}
  Analogous to figure \ref{Fig:L2_carpet_Euler_angle} the distribution
  of $\rho(\hat n)$ is shown for the torus ensemble with $L=3$.}
\end{figure}

The results are shown in figures \ref{Fig:L2_carpet_Euler_angle} and
\ref{Fig:L3_carpet_Euler_angle} for the torus ensembles with side-lengths
$L=2$ and $L=3$, respectively.
The analysis in \cite{Aurich_Buchert_France_Steiner_2021} corresponds
to the lower left corner of the figures at
$\Phi_{\textrm{\scriptsize or}} = \Theta_{\textrm{\scriptsize or}} = 0$.
One receives from the colour bars a variation of $\Delta\rho \approx 0.25$,
which is small in comparison to the change of the mean value from
39.7 to 36.2 for $L=2$ to $L=3$.
Therefore, the violation of isotropy would not change the conclusion
of \cite{Aurich_Buchert_France_Steiner_2021}.
A detailed analysis is given in the companion paper
\cite{Aurich_France_Steiner_2026}.

\begin{figure}
\begin{center}
\vspace*{0pt}
\begin{minipage}{15.8cm}
\hspace*{0pt}\includegraphics[width=16.0cm]{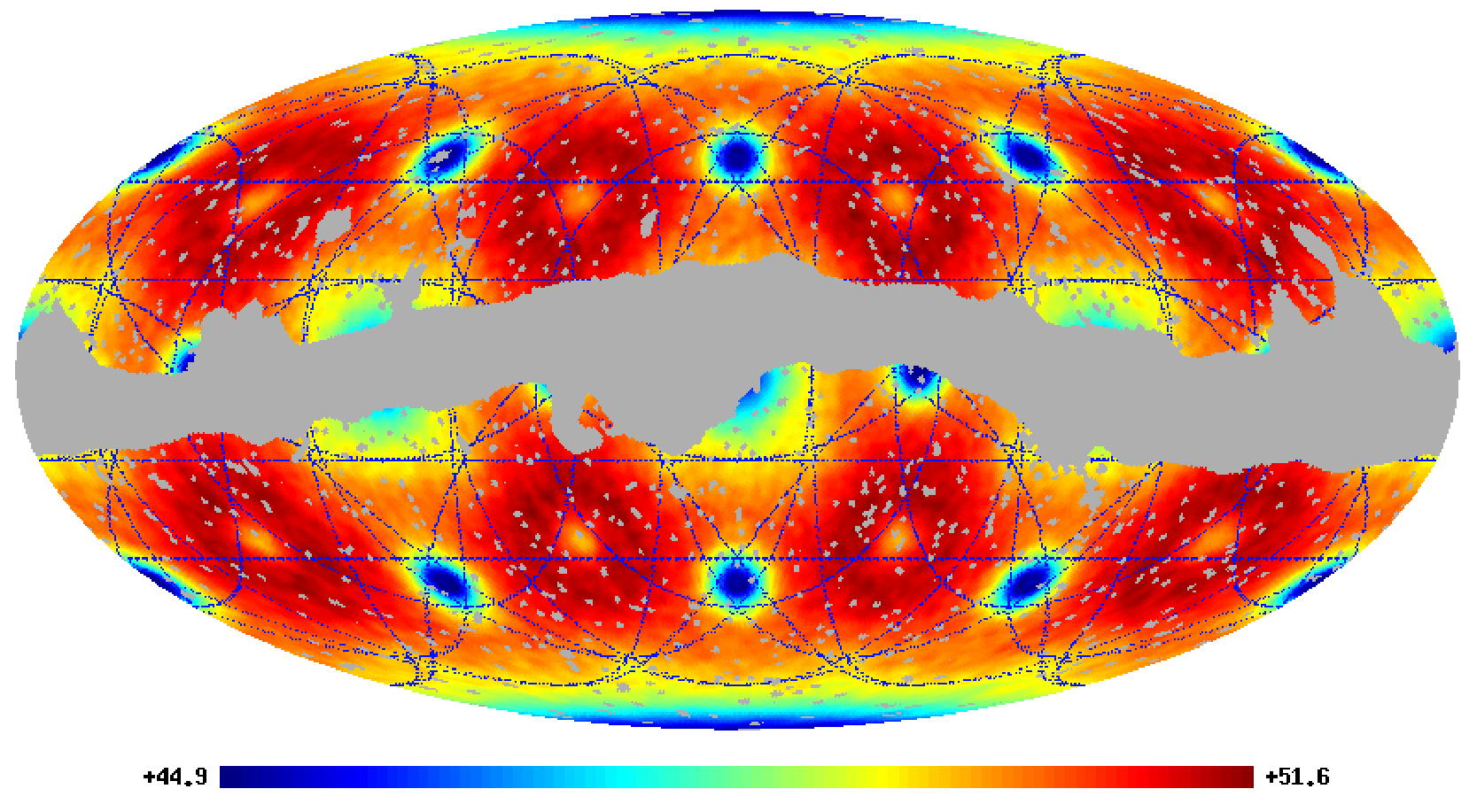}\vspace*{20pt}
\hspace*{0pt}\includegraphics[width=16.0cm]{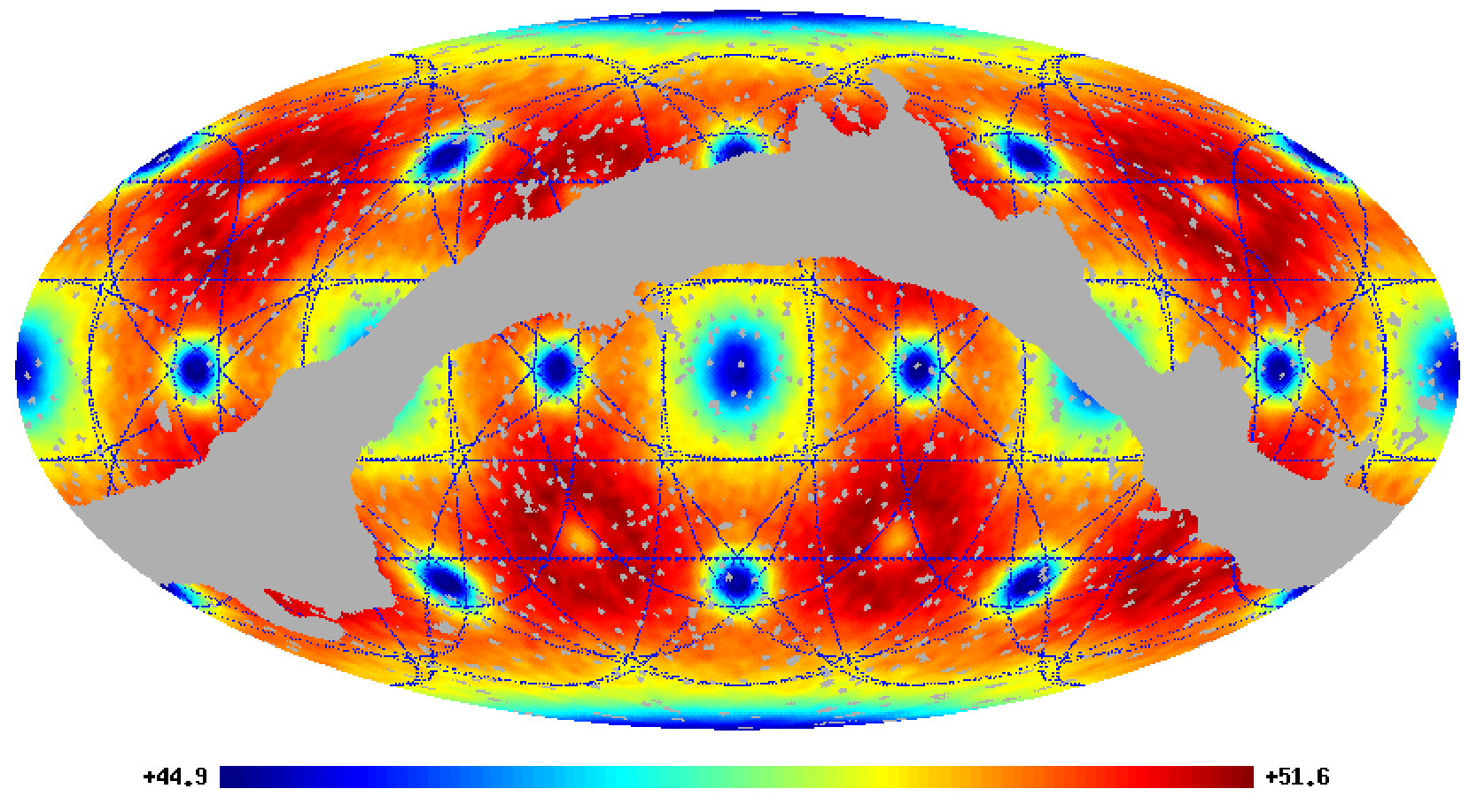}
\end{minipage}
\put(-445,266){$(a)$ mask orientation for the minimum in $\rho$
\hspace{80pt} $L=2$}
\put(-445,-5){$(b)$ mask orientation for the maximum in $\rho$
\hspace{80pt} $L=2$}
\end{center}
\vspace*{0pt}
\caption{\label{Fig:L2_StdDev_mask_min_max}
  These 2 panels (a above and b below) explain the relative distribution
  of the extrema in $\rho$ shown in figure \ref{Fig:L2_carpet_Euler_angle} with $L$=2.
  The average of the standard deviation $\sigma(\hat n)$ computed
  from 100\,000 torus simulations with side-length $L=2$ is shown
  together with the orientation of the mask that leads to the minimum
  (panel (a)) in $\rho$ and to the maximum (panel (b)).
  For the mask orientation leading to a minimum in $\rho$,
  the mask eliminates 8 minima in $\sigma(\hat n)$ (blue domains),
  while for the orientation giving a maximum in $\rho$
  only 4 minima are masked.
}
\end{figure}

\begin{figure}
\begin{center}
\vspace*{0pt}
\begin{minipage}{15.8cm}
\hspace*{0pt}\includegraphics[width=16.0cm]{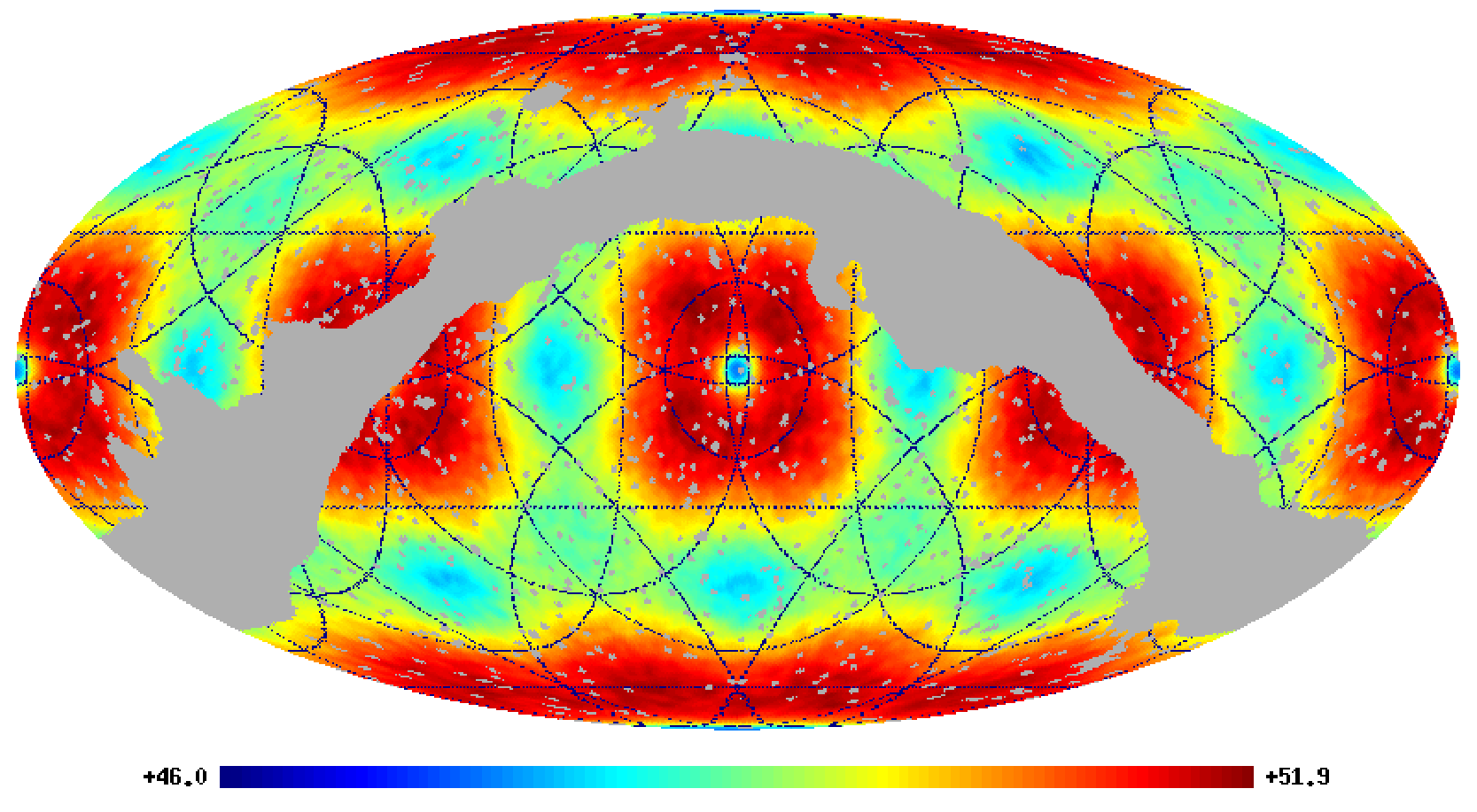}\vspace*{20pt}
\hspace*{0pt}\includegraphics[width=16.0cm,]{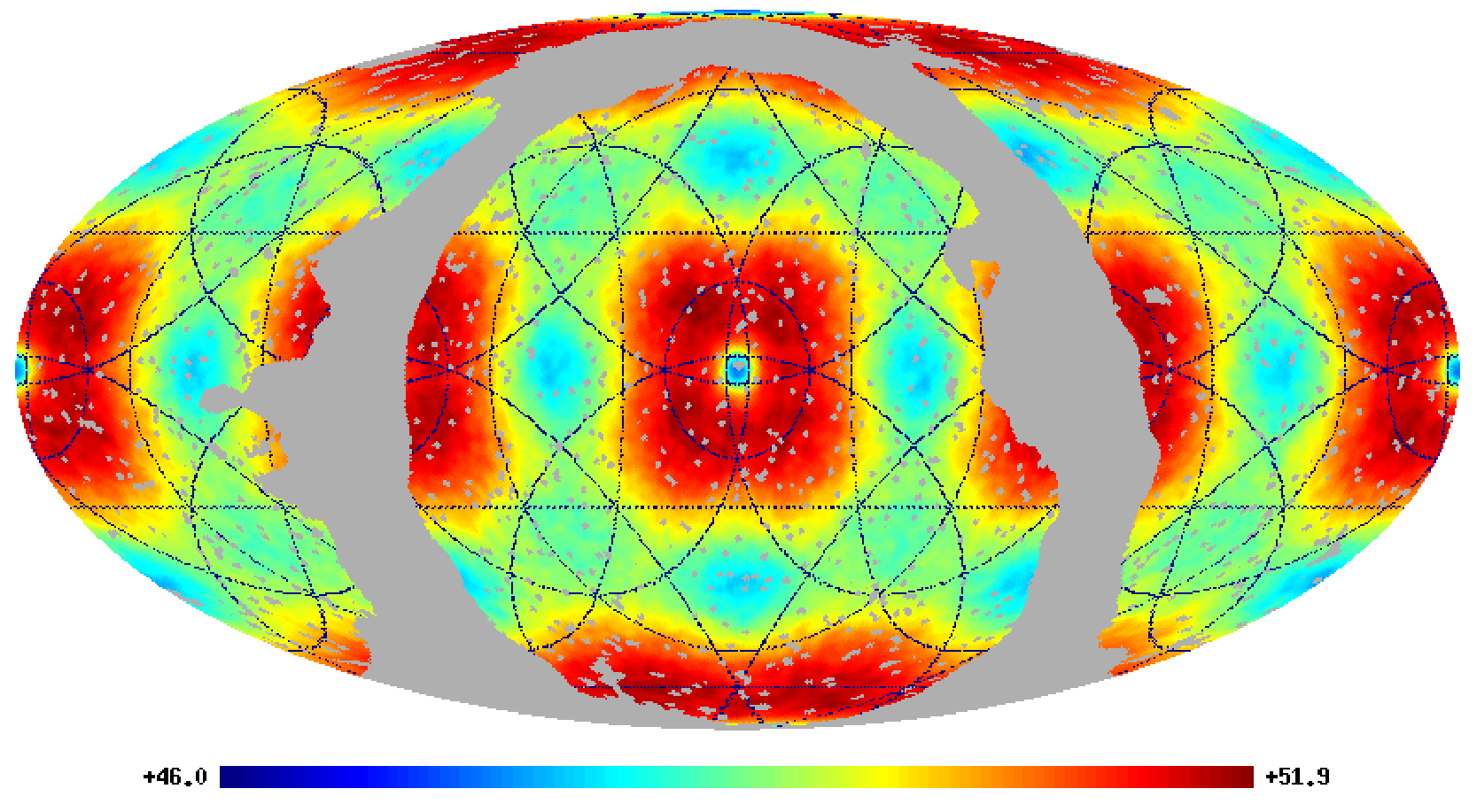}
\end{minipage}
\put(-445,266){$(a)$ mask orientation for the minimum in $\rho$
\hspace{80pt} $L=3$}
\put(-445,-5){$(b)$ mask orientation for the maximum in $\rho$
\hspace{80pt} $L=3$}
\end{center}
\vspace*{0pt}
\caption{\label{Fig:L3_StdDev_mask_min_max}
  The same as in figure \ref{Fig:L2_StdDev_mask_min_max}
  but for the case $L=3$.
  For the mask orientation leading to a minimum in $\rho$,
  the mask eliminates only 2 maxima in $\sigma(\hat n)$ (red domains),
  while for the orientation providing a maximum in $\rho$,
  4 maxima in $\sigma(\hat n)$ are masked.
}
\end{figure}

As discussed in the context of figures \ref{Fig:L2_StdDev} and
\ref{Fig:L2_Square_Derivative},
the violation of the isotropy is more pronounced in
the standard deviation $\sigma(\hat n)$ than
in the gradient quantity $\sigma_1(\hat n)$.
Since $\sigma(\hat n)$ is the denominator of $\rho(\hat n)$,
see equation (\ref{def_rho_directional}),
regions with low values in $\sigma(\hat n)$ correspond to regions
with large values in $\rho(\hat n)$ and conversely.
In figures \ref{Fig:L2_StdDev_mask_min_max} and
\ref{Fig:L3_StdDev_mask_min_max},
the orientation of the mask is shown for the maximal value of $\rho(\hat n)$
and for the minimal value obtained from the calculation shown
in figures \ref{Fig:L2_carpet_Euler_angle} and
\ref{Fig:L3_carpet_Euler_angle} for $L=2$ and $L=3$, respectively.
This nicely illustrates that the variation in $\rho(\hat n)$ is explained 
by the anisotropy pattern of the standard deviation $\sigma(\hat n)$.
Large values in $\rho(\hat n)$ occur where the mask obscures more regions
with large values of $\sigma(\hat n)$ and less regions with small values.
Explicit examples are discussed in the figure captions of figures
\ref{Fig:L2_StdDev_mask_min_max} and \ref{Fig:L3_StdDev_mask_min_max}.

\section{Summary and Discussion}

\label{summary}


This paper ponders the question whether one of the simplest topological spaces,
the cubic 3-torus in flat space,
induces sufficient statistical anisotropy over the CMB sky
such that it can be detected.
It turns out that very large 3-torus ensembles are needed
in order to reveal the tiny deviations from the statistical isotropy
in the 3-torus CMB simulations.
Even an ensemble size of 100\,000 members does not reveal a
deviation in the position dependent mean value of the temperature
field $\delta T(\hat n)$,
only its variance $\sigma^2(\hat n)$ displays a minute
position dependence as exemplarily revealed in
figure \ref{Fig:Histogram_max_min_comparison}.
This shows that the statistical anisotropy is determined by a
distribution of $\delta T(\hat n)$
with varying variance but having the same mean value.
The next higher statistics $\sigma_1^2(\hat n)$ constructed from the
covariant derivative of the temperature field $\delta T(\hat n)$
does not display a pronounced anisotropy as the variance $\sigma^2(\hat n)$,
as a comparison of figure \ref{Fig:L2_StdDev_withDip} with figure
\ref{Fig:L2_Square_Derivative} reveals.

Orbit sum rules are used in order to highlight the observed patterns
of statistical anisotropy in the position dependence
of the variance $\sigma^2(\hat n)$.
To this aim the Sachs-Wolfe formula is adapted such that the sum over
the eigenmodes of the 3-torus in the variance $\sigma^2(\hat n)$
can be replaced by a sum over the lengths
$d(\vec x(\eta_{\hbox{\scriptsize{SLS}}}),  \gamma(\vec 0\,))$ of orbits.
The orbits occurring in the sum rules connect the given point
$\vec x(\eta_{\hbox{\scriptsize{SLS}}})$ on the surface of last scattering
with the topological mirror points $\gamma(\vec 0\,)$ of the CMB observer
at $\vec 0$, where the dominant contribution arises from short distances.
For a fixed group element $\gamma$,
the shortest distance does not occur along the circles-in-the-sky,
but instead at its midpoint as illustrated by figures
\ref{Fig:L2_Generatoren} and \ref{Fig:L_2_Zweierprodukte}.
It is the position of minimal distance of a mirror observer
at $\gamma(\vec 0\,)$ to the surface of last scattering.
Thus positions on the surface of last scattering,
which are such midpoints, are prone to extreme values in
the variance $\sigma^2(\hat n)$.
The group elements $\gamma\in\Gamma$ which correspond to directly neighbouring
copies of the fundamental cell have thus the greatest influence.
This is again demonstrated by figure \ref{Fig:L2_Generatoren},
where only the circles due to the generators of $\Gamma$ are shown,
and by figure \ref{Fig:L_2_Zweierprodukte}, which also shows circles
due to a product of two generators.
Since the path between the mirror observer at $\gamma(\vec 0\,)$ and
the surface of last scattering with minimal distance intersects the
surface perpendicularly and, of course, the observer at $\vec 0$ sits
in the centre of this spherical surface, the path is part of the periodic orbit,
which connects the observer position at $\vec 0$ with its mirror point
at $\gamma(\vec 0\,)$.
In this way, periodic orbits belonging to the group $\Gamma$ determine
the violation of statistical isotropy.

Although the deviations from the statistical isotropy are small,
it is interesting to consider the normalised modulus $\rho$
of the gradient of the temperature field $\delta T(\hat n)$,
because the standard deviation $\sigma(\hat n)$ occurs in the denominator
of $\rho$.
It is shown that in the case of a masked sky,
the anisotropy depends slightly on the orientation of a mask
due to obscuring a different number of regions with extreme values in 
$\sigma(\hat n)$.
See the impact over $\rho$, in figures \ref{Fig:L2_carpet_Euler_angle}
and \ref{Fig:L3_carpet_Euler_angle},
of the mask position hiding $\sigma$ regions in
figures \ref{Fig:L2_StdDev_mask_min_max} and
\ref{Fig:L3_StdDev_mask_min_max}.

The study of this paper addresses only the 3-torus $\mathbb{T}^3$,
which is special in the sense that it is globally homogeneous.
Thus it is possible that other flat topologies do not reveal the
extremely faint breaking of the statistical isotropy of
the 3-torus $\mathbb{T}^3$,
but this is left to future investigations.


\appendix

\section{Derivation of the orbit sum rule on the cubic 3-torus $\mathbb{T}^3$}

\setcounter{equation}{0}
\renewcommand{\theequation}{A.\arabic{equation}}

Starting point is the following relation for the weighted correlation
function of the eigenfunctions $\psi_{\vec \alpha}(\vec x\,)$
(see \cite{Aurich_Steiner_1991,Aurich_Steiner_1995,Aurich_Steiner_2001})
\begin{eqnarray} \label{Green_sum}
  C_\Gamma(\vec x,\vec x\,') & := &
  \sum_{\vec\alpha} F(E_{\vec\alpha}) \,
  \psi_{\vec \alpha}(\vec x\,) \, \psi^\star_{\vec \alpha}(\vec x\,')
  \\ & = & \label{Green_integral}
  \frac 1\pi \int_0^\infty F(E) \,
  \hbox{Im}\, G_\Gamma(\vec x,\vec x\,'; E+i\epsilon)\, dE
  \hspace{10pt} ,
\end{eqnarray}
where $G_\Gamma(\vec x,\vec x\,'; E)$ is the Green function of the cubic torus,
$(\Delta_{\vec x}+E)\,G_\Gamma(\vec x,\vec x\,'; E) = \delta^3(\vec x-\vec x\,')$
with $\vec x,\vec x\,'\in\mathbb{T}^3$,
and the weight function $F(E)$ is assumed to be holomorphic in a strip enclosing
the real axis $E\geq0$.

The imaginary part in (\ref{Green_integral}) is defined as the discontinuity
of $G_\Gamma$ ($\epsilon > 0$)
\begin{equation}
  \hbox{Im}\, G_\Gamma(\vec x,\vec x\,'; E+i\epsilon) \; = \;
  \frac 1{2i} \left[ G_\Gamma(\vec x,\vec x\,'; E+i\epsilon) -
    G_\Gamma(\vec x,\vec x\,'; E-i\epsilon) \right]
  \hspace{3pt} .
\end{equation}
In order that the sum in (\ref{Green_sum}),
i.\,e.\ the sum over the eigenvalues $E_{\vec\alpha} = (2\pi/L)^2 \vec\alpha^2$
of $\mathbb{T}^3$, is absolutely convergent
$$
|C_\Gamma(\vec x,\vec x\,')| \; \leq \; \frac 1V \sum_{\vec\alpha} |F(E_{\vec\alpha})|
\; < \; \infty\; \; ,
$$
$F(E)$ has to decrease faster than $1/E^{3/2}$ for $E\to\infty$.\\
\underline{Proof:}\\
Consider the general sum $(s\in\mathbb{C})$
$$
\sum_{\vec\alpha} {'} \frac 1{(E_{\vec\alpha})^{s/2}} \; = \;
\left(\frac L{2\pi}\right)^s \, \sum_{\vec\alpha} {'} \frac 1{|\vec\alpha|^s}
\; =: \;
\left(\frac L{2\pi}\right)^s \, \zeta_E(s) \hspace{10pt} ,
$$
where $\zeta_E(s)$ denotes the Epstein zeta function
\cite{Epstein_1903,Epstein_1906} (see also \cite{Steiner_2016}).  
Epstein proved that the Dirichlet series of $\zeta_E(s)$ converges absolutely
for $\hbox{Re}s>3$ (= dimension of $\mathbb{T}^3$) and diverges for
$\hbox{Re}s\leq 3$. \hfill q.e.d.\\
Note, however, that $\zeta_E(s)$ has an analytic continuation as a
meromorphic function to the whole complex $s$-plane with a single pole at $s=3$
(similar to the Riemann zeta function having a pole at $s=1$).

Based on the method of images, the torus Green function $G_\Gamma$ is written
as a Poincar\'e series over the group elements $\gamma$ of
the torus group $\Gamma$ as
\begin{equation}
  G_\Gamma(\vec x,\vec x\,'; E+i\epsilon) \; = \;
  \sum_\gamma G_0(\vec x,\gamma(\vec x\,'); E+i\epsilon)
  \hspace{10pt} ,
\end{equation}
where $G_0$ denotes the Green function on the covering manifold $\mathbb{E}^3$.
With the well-known expression $(\epsilon>0)$
$$
G_0(\vec x,\vec x\,'; E+i\epsilon) \; = \;
\frac 1{4\pi} \, \frac 1{|\vec x-\vec x\,'|} \,
e^{-|\vec x-\vec x\,'| \sqrt{-(E+i\epsilon)}}
$$
one obtains
\begin{equation}
  \lim_{\epsilon\to 0^+} \hbox{Im}\, G_0(\vec x,\vec x\,'; E+i\epsilon) \; = \;
  \frac 1{4\pi} \, \frac{\sin(k|\vec x-\vec x\,'|)}{|\vec x-\vec x\,'|}
\end{equation}
with $k = \sqrt{E}$.
Redefining the spectral function
\begin{equation}
h(k) \; := \; F(E) \; = \; F(k^2)
\end{equation}
obeying 
\begin{itemize}
\item[(i)] $h(k) \; = \; h(-k)$
\item[(ii)] $h(k)$ is holomorphic in a strip of the positive $k$ axis
  with $\epsilon \leq k \leq\infty$, $\epsilon>0$
\item[(iii)] $h(k) = O(1/(|k|^{3+\delta}))$ for $|k|\to \infty$,
  $\delta>0$ \hspace{10pt} ,
\end{itemize}
we obtain the orbit sum rule
\begin{eqnarray}
  C_\Gamma(\vec x,\vec x\,') & := & \nonumber
  \sum_{\vec\alpha} h(k_{\vec\alpha}) \,
  \psi_{\vec \alpha}(\vec x\,) \, \psi^\star_{\vec \alpha}(\vec x\,')
  \\  & = &
  \frac 1{2\pi^2} \, \sum_\gamma \frac{\widetilde{h}(d(\vec x,\gamma(\vec x\,'))}
                                    {d(\vec x,\gamma(\vec x\,'))}
\end{eqnarray}
with 
\begin{equation}
\widetilde{h}(y) \; := \; \int_0^\infty \, k \, \sin(k\,y) \, h(k) \, dk
\hspace{10pt} ,
\end{equation}
where $d(\vec x,\gamma(\vec x\,'))$ denotes the Euclidean distance of the
point $\vec x$ to the mirror point $\gamma(\vec x\,')$ of $\vec x\,'$ due
to the group element $\gamma\in\Gamma$.

The spectral function
\begin{equation}\label{appendix_fkt2_heat_kernel}
  h(k) \; = \; \frac 1{k^2}e^{-k^2/(2\sigma^2)}
\end{equation}
is needed in the main text.
In this case, the corresponding function $\widetilde{h}(y)$ has to be
regularised by subtracting the zero-mode explicitly $(y>0)$
\begin{eqnarray}
  \widetilde{h}(y) & = &
  \int_0^\infty \, k \, \sin(k\,y) \, \frac 1{k^2}e^{-k^2/(2\sigma^2)} \, dk
  \\ \nonumber & = &
  \underbrace{\int_0^\infty \, \frac{\sin(k\,y)}k \, dk}_{= \pi/2} \, + \,
  \int_0^\infty \, k \, \sin(k\,y) \,
  \underbrace{\Big[\frac 1{k^2}e^{-k^2/(2\sigma^2)} - \frac 1{k^2} \Big]}_
             {= h_{\hbox{\scriptsize{reg}}}(k)}\, dk
  \\ \nonumber & = &
  \frac\pi 2 \, + \, \widetilde{h}_{\hbox{\scriptsize{reg}}}(y)      
  \hspace{10pt} .
\end{eqnarray}
Then $\widetilde{h}_{\hbox{\scriptsize{reg}}}(y)$ has no contribution due
to the zero mode and thus
\begin{equation}\label{appendix_fkt2_heat_kernel_tilde}
  \widetilde{h}_{\hbox{\scriptsize{reg}}}(y) \; = \;
  -\;\frac{\pi}2\; \hbox{erfc}\left(\frac{\sigma}{\sqrt{2}}\,y\right)
  \hspace{10pt} .
\end{equation}


\section*{Acknowledgements}

The software packages HEALPix (\url{http://healpix.sourceforge.io},
\cite{Gorski_Hivon_Banday_Wandelt_Hansen_Reinecke_Bartelmann_2005})
and CAMB written by A.~Lewis and A.~Challinor (\url{http://camb.info})
as well as the Planck data from 
\url{http://pla.esac.esa.int/pla/#maps} and
\url{https://irsa.ipac.caltech.edu/data/Planck/} were used in this work.


\section*{References}

\bibliography{../bib_astro}

\bibliographystyle{iopart-num-doi}

\label{lastpage}

\end{document}